\documentclass[longauth]{aa}  

\usepackage{graphicx}
\usepackage{txfonts}
\usepackage{url}
\usepackage[colorlinks=true,linkcolor=blue]{hyperref}
\hypersetup{
    colorlinks,
    linkcolor={red!50!black},
    citecolor={blue!50!black},
    urlcolor={blue!80!black}
}
\usepackage{adjustbox}
\usepackage{blindtext}
\usepackage{amsmath}
\usepackage[switch]{lineno} 
\usepackage{svg}
\usepackage{xcolor}
\usepackage{siunitx}
\usepackage{enumitem}
\usepackage{booktabs}
\usepackage{natbib}
\usepackage{float}

\newcommand{\orcid}[1]{\textsuperscript{\href{https://orcid.org/#1}{\includegraphics[width=8pt]{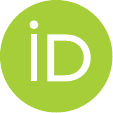}}}}

\newcommand{\fsig}{f{\sigma_8}}

\newcommand{\hmpc}{\,h^{-1}\,\text{Mpc}}

\begin{document}

   \title{The DESI DR1 Peculiar Velocity Survey: Mock Catalogs}


\authorrunning{J.~Bautista et al.}
\author{
J.~Bautista \inst{1},
A.~J.~Amsellem \inst{2},
V.~Aronica \inst{1},
S.~BenZvi \inst{3},
C.~Blake \inst{4},
A.~Carr \inst{5},
T.~M.~Davis \inst{6},
K.~Douglass \inst{3},
T.~Dumerchat \inst{1,7},
C.~Howlett \inst{6},
Y.~Lai \inst{6,8},
A.~Nguyen \inst{4},
A.~Palmese \inst{2},
F.~Qin \inst{1},
C.~Ravoux \inst{9},
C.~Ross \inst{6},
K.~Said \inst{6},
R. J.~Turner \inst{4},
J.~Aguilar \inst{10},
S.~Ahlen \inst{11},
D.~Bianchi \inst{12,13},
D.~Brooks \inst{14},
T.~Claybaugh \inst{10},
A.~Cuceu \inst{10},
A.~de la Macorra \inst{15},
P.~Doel \inst{14},
A.~Font-Ribera \inst{7},
J.~E.~Forero-Romero \inst{16,17},
E.~Gaztañaga \inst{18,19,20},
S.~Gontcho A Gontcho \inst{10,21},
G.~Gutierrez \inst{22},
H.~K.~Herrera-Alcantar \inst{23,24},
K.~Honscheid \inst{25,26,27},
D.~Huterer \inst{28,29},
M.~Ishak \inst{30},
R.~Joyce \inst{31},
A.~Kremin \inst{10},
C.~Lamman \inst{27},
M.~Landriau \inst{10},
L.~Le~Guillou \inst{32},
A.~Leauthaud \inst{33,34},
M.~Manera \inst{35,7},
A.~Meisner \inst{31},
R.~Miquel \inst{36,7},
J.~Moustakas \inst{37},
A.~Muñoz-Gutiérrez \inst{15},
S.~Nadathur \inst{19},
W.~J.~Percival \inst{38,39,40},
F.~Prada \inst{41},
I.~P\'erez-R\`afols \inst{42},
G.~Rossi \inst{43},
E.~Sanchez \inst{44},
D.~Schlegel \inst{10},
M.~Schubnell \inst{28,29},
H.~Seo \inst{45},
J.~Silber \inst{10},
D.~Sprayberry \inst{31},
G.~Tarl\'{e} \inst{29},
B.~A.~Weaver \inst{31},
P.~Zarrouk \inst{32},
R.~Zhou \inst{10},
and H.~Zou \inst{46},
}

\institute{
Aix Marseille Univ, CNRS/IN2P3, CPPM, Marseille, France 
\and Department of Physics, Carnegie Mellon University, 5000 Forbes Avenue, Pittsburgh, PA 15213, USA 
\and Department of Physics \& Astronomy, University of Rochester, 206 Bausch and Lomb Hall, P.O. Box 270171, Rochester, NY 14627-0171, USA 
\and Centre for Astrophysics \& Supercomputing, Swinburne University of Technology, P.O. Box 218, Hawthorn, VIC 3122, Australia 
\and Korea Astronomy and Space Science Institute, 776, Daedeokdae-ro, Yuseong-gu, Daejeon 34055, Republic of Korea 
\and School of Mathematics and Physics, University of Queensland, Brisbane, QLD 4072, Australia 
\and Institut de F\'{i}sica d’Altes Energies (IFAE), The Barcelona Institute of Science and Technology, Edifici Cn, Campus UAB, 08193, Bellaterra (Barcelona), Spain 
\and Steward Observatory, University of Arizona, 933 N. Cherry Avenue, Tucson, AZ 85721, USA 
\and Universit\'{e} Clermont-Auvergne, CNRS, LPCA, 63000 Clermont-Ferrand, France 
\and Lawrence Berkeley National Laboratory, 1 Cyclotron Road, Berkeley, CA 94720, USA 
\and Department of Physics, Boston University, 590 Commonwealth Avenue, Boston, MA 02215 USA 
\and Dipartimento di Fisica ``Aldo Pontremoli'', Universit\`a degli Studi di Milano, Via Celoria 16, I-20133 Milano, Italy 
\and INAF-Osservatorio Astronomico di Brera, Via Brera 28, 20122 Milano, Italy 
\and Department of Physics \& Astronomy, University College London, Gower Street, London, WC1E 6BT, UK 
\and Instituto de F\'{\i}sica, Universidad Nacional Aut\'{o}noma de M\'{e}xico,  Circuito de la Investigaci\'{o}n Cient\'{\i}fica, Ciudad Universitaria, Cd. de M\'{e}xico  C.~P.~04510,  M\'{e}xico 
\and Departamento de F\'isica, Universidad de los Andes, Cra. 1 No. 18A-10, Edificio Ip, CP 111711, Bogot\'a, Colombia 
\and Observatorio Astron\'omico, Universidad de los Andes, Cra. 1 No. 18A-10, Edificio H, CP 111711 Bogot\'a, Colombia 
\and Institut d'Estudis Espacials de Catalunya (IEEC), c/ Esteve Terradas 1, Edifici RDIT, Campus PMT-UPC, 08860 Castelldefels, Spain 
\and Institute of Cosmology and Gravitation, University of Portsmouth, Dennis Sciama Building, Portsmouth, PO1 3FX, UK 
\and Institute of Space Sciences, ICE-CSIC, Campus UAB, Carrer de Can Magrans s/n, 08913 Bellaterra, Barcelona, Spain 
\and University of Virginia, Department of Astronomy, Charlottesville, VA 22904, USA 
\and Fermi National Accelerator Laboratory, PO Box 500, Batavia, IL 60510, USA 
\and Institut d'Astrophysique de Paris. 98 bis boulevard Arago. 75014 Paris, France 
\and IRFU, CEA, Universit\'{e} Paris-Saclay, F-91191 Gif-sur-Yvette, France 
\and Center for Cosmology and AstroParticle Physics, The Ohio State University, 191 West Woodruff Avenue, Columbus, OH 43210, USA 
\and Department of Physics, The Ohio State University, 191 West Woodruff Avenue, Columbus, OH 43210, USA 
\and The Ohio State University, Columbus, 43210 OH, USA 
\and Department of Physics, University of Michigan, 450 Church Street, Ann Arbor, MI 48109, USA 
\and University of Michigan, 500 S. State Street, Ann Arbor, MI 48109, USA 
\and Department of Physics, The University of Texas at Dallas, 800 W. Campbell Rd., Richardson, TX 75080, USA 
\and NSF NOIRLab, 950 N. Cherry Ave., Tucson, AZ 85719, USA 
\and Sorbonne Universit\'{e}, CNRS/IN2P3, Laboratoire de Physique Nucl\'{e}aire et de Hautes Energies (LPNHE), FR-75005 Paris, France 
\and Department of Astronomy and Astrophysics, UCO/Lick Observatory, University of California, 1156 High Street, Santa Cruz, CA 95064, USA 
\and Department of Astronomy and Astrophysics, University of California, Santa Cruz, 1156 High Street, Santa Cruz, CA 95065, USA 
\and Departament de F\'{i}sica, Serra H\'{u}nter, Universitat Aut\`{o}noma de Barcelona, 08193 Bellaterra (Barcelona), Spain 
\and Instituci\'{o} Catalana de Recerca i Estudis Avan\c{c}ats, Passeig de Llu\'{\i}s Companys, 23, 08010 Barcelona, Spain 
\and Department of Physics and Astronomy, Siena University, 515 Loudon Road, Loudonville, NY 12211, USA 
\and Department of Physics and Astronomy, University of Waterloo, 200 University Ave W, Waterloo, ON N2L 3G1, Canada 
\and Perimeter Institute for Theoretical Physics, 31 Caroline St. North, Waterloo, ON N2L 2Y5, Canada 
\and Waterloo Centre for Astrophysics, University of Waterloo, 200 University Ave W, Waterloo, ON N2L 3G1, Canada 
\and Instituto de Astrof\'{i}sica de Andaluc\'{i}a (CSIC), Glorieta de la Astronom\'{i}a, s/n, E-18008 Granada, Spain 
\and Departament de F\'isica, EEBE, Universitat Polit\`ecnica de Catalunya, c/Eduard Maristany 10, 08930 Barcelona, Spain 
\and Department of Physics and Astronomy, Sejong University, 209 Neungdong-ro, Gwangjin-gu, Seoul 05006, Republic of Korea 
\and CIEMAT, Avenida Complutense 40, E-28040 Madrid, Spain 
\and Department of Physics \& Astronomy, Ohio University, 139 University Terrace, Athens, OH 45701, USA 
\and National Astronomical Observatories, Chinese Academy of Sciences, A20 Datun Road, Chaoyang District, Beijing, 100101, P.~R.~China 
}

   \date{Received ; accepted }

 
    \abstract
    {We describe the production of the official set of mock catalogs for the Dark Energy Spectroscopic Instrument Peculiar Velocity Survey (DESI-PV) Data Release 1 (DR1).  
    Our mock catalogs reproduce the Bright Galaxy Survey 
    number density and clustering at low redshift $(z<0.1)$
    and the DESI PV samples of Fundamental plane and Tully-Fisher distances, from which we derive peculiar velocities. 
    We carefully match mock and data properties
    and we mimic measurements of distance indicators and peculiar velocities, 
    which follow the same statistical properties as real data. 
    Mock samples of type-Ia supernovae also complement the other two distance indicators. 
    Our 675 available mock realizations were used consistently by our three different methodologies described in our companion papers
    that measure the growth rate of structure $f
    \sigma_8$ with DESI PV DR1. 
    Those mocks allow us to perform precise tests of clustering models and uncertainty estimation to an unprecedented level of accuracy, and compute correlations between 
    methodologies. The consensus value for the DESI DR1 PV growth rate measurement is $f\sigma_8 = 0.450 \pm 0.055$. 
    This sample of mock catalogs represents the largest and most realistic set for cosmological measurements with peculiar velocities to date. 
    }

    \keywords{cosmology -- 
                peculiar velocities --
                growth-rate of structures --
                spectroscopic surveys
               }

    \maketitle
%

\section{Introduction}

 {Measuring the statistical properties of cosmological density and velocity fields on scales of tens of Megaparsecs (Mpc) teaches us about the nature and abundances of dark matter and dark energy in our Universe. Those properties also depend on important cosmological parameters such as the Hubble constant $H_0$ or the variance of linear matter density perturbations smoothed over spheres of $8~\hmpc$, $\sigma_8$.  }
The large-scale correlations of density and velocity fields can also be used to test our current description of gravity: general relativity (GR). Measurements of the growth rate of structures as a function of cosmic time can be obtained from such fields, and be compared to GR predictions. 
In GR, the growth rate of structures can be written as $f(z) \approx [\Omega_m(z)]^\gamma$, where $\Omega_m(z)$ is the cosmic fraction of cold dark + baryonic matter as a function of redshift $z$ and $\gamma = 0.55$ is the gravitational growth index \citep{linderParameterizedBeyondEinsteinGrowth2007}. 
A variety of alternatives/modifications to GR have been proposed to explain the accelerated nature of the expansion without dark energy (see \citealt{ferreiraCosmologicalTestsGravity2019,ishakTestingGeneralRelativity2019,houCosmologicalProbesStructure2023,hutererGrowthCosmicStructure2023} 
for reviews on the topic).  Some modified gravity models predict the same  expansion history as dark energy models, however, they predict a quite different growth rate history. Precise measurements of the growth rate combined with the expansion rate have the potential to uncover departures from GR \citep{ishakModifiedGravityConstraints2025}. 

At redshifts $z>0.1$, growth rate measurements have been typically obtained from the anisotropies observed in the two-point functions of spectroscopic galaxy surveys. Such anisotropies arise from the impact of peculiar velocities on the observed redshifts, creating distortions on the observed density field since we cannot discriminate between Hubble flow and proper motions. In the past twenty years, many surveys used redshift-space distortions (RSD) to measure growth rates, including 
including 
   WiggleZ \citep{blakeWiggleZDarkEnergy2011a}, 
   6dFGRS \citep{beutler6dFGalaxySurvey2012}, 
   SDSS-II \citep{samushiaInterpretingLargescaleRedshiftspace2012}, 
   SDSS-MGS \citep{howlettClusteringSDSSMain2015},
   FastSound \citep{okumuraSubaruFMOSGalaxy2016},
   VIPERS \citep{pezzottaVIMOSPublicExtragalactic2017, delatorreVIMOSPublicExtragalactic2017}, 
   SDSS-III BOSS \citep{beutlerClusteringGalaxiesCompleted2017, griebClusteringGalaxiesCompleted2017, sanchezClusteringGalaxiesCompleted2017, satpathyClusteringGalaxiesCompleted2017},
   SDSS-IV eBOSS 
   \citep{bautistaCompletedSDSSIVExtended2021, 
   gil-marinCompletedSDSSIVExtended2020,
   demattiaCompletedSDSSIVExtended2021, 
   tamoneCompletedSDSSIVExtended2020, 
   houCompletedSDSSIVExtended2021, 
   neveuxCompletedSDSSIVExtended2020}, 
   and finally the Dark Energy Spectroscopic Instrument (DESI) 
   \citep{desicollaborationDESI2024FullShape2025}.

At redshifts $z< 0.1$, it is possible to estimate individual peculiar velocities of galaxies by means of distance indicators, such as the Tully-Fisher relation for spiral galaxies \citep{tullyNewMethodDetermining1977}, the Fundamental plane for ellipticals \citep{djorgovskiFundamentalPropertiesElliptical1987} and type-Ia supernovae \citep{kowalAbsoluteMagnitudesSupernovae1968}. Such indicators break the degeneracy between Hubble flow and proper motions, provided an observed redshift. With a large sample of galaxy distances and redshifts, a radial peculiar velocity field can be constructed and their large-scale correlations measured. Those correlations are directly proportional to the growth rate of structures, so they become competitive in a regime where RSD lacks volume for precise measurements. 
Therefore, a joint analysis of the spatial correlations of the redshift-space density and radial velocity fields, including their cross-correlations,  yields more precise estimates of the growth rate. 
That is the general context of this work. 

Many methodologies have been developed in the literature 
that perform a joint analysis of density and velocity fields 
to measure the growth rate of structures. 
The first one consists of compressing the field into 
two-point statistics as a function of scale. 
Since we have two fields, we can estimate three two-point functions: 
density-density, density-velocity and velocity-velocity. 
Those $3\times2$-point functions can be measured and modeled either in configuration space 
\citep{Gorski1989,Howlett2015,Adams2017,Dupuy2019,YuyuWang2018,YuyuWang2021,Turner2021,qinHIHODHalo2022,qinGalaxyNumberDensity2023,Turner2023} 
or in Fourier space \citep{Feldman1994,Park2000,Park2006,Yamamoto2006,Blake2010,Zhang2017,Howlett2019,Qin2019b,Appleby2023,Shi2024,Qin2025}. 
A second method consists of working at the fields level, modeling the covariance between positions in the volume in a large likelihood, assuming those fields are a random realization of a multivariate Gaussian
 \citep{Johnson2014,Howlett2017,Adams2020,Lai2023,Carreres2023}.  
A third type of method compares the measured velocities to reconstructed velocities from the density field \citep{Pike2005,Erdogdu2006,Ma2012,Springob2014,Carrick2015,Said2020,Qin2023,Lilow2024,boubelLargescaleMotionsGrowth2024,hollingerCosmologicalParametersEstimated2024a}.
Each methodology relies on different sets of assumptions and approximations,
so final measurements might differ. In this work, we produced a large 
set of simulated datasets enabling us, for the first time, to test
consistently these three different methodologies to the highest level of precision 
\citep{DESIPV_Lai, DESIPV_Qin, DESIPV_Turner}. 

In this work, we present the official set of mock catalogs from the DESI Peculiar Velocity Survey Data Release 1. 
Such mocks were essential for 
 (1) validating the pipeline of the data production; 
 (2) exploring how selection functions and distance calibrations affect the measurements; 
 (3) estimating the covariance matrices of the measured observables; 
 and (4) validating the methods/estimators used to extract the cosmological parameters. 
 These tasks require advanced mock sampling algorithms which can accurately reproduce the selection functions, survey geometry, galaxy and velocity clustering of the real data. 
  {The sample presented in this work is composed of 675 realizations based on n-body simulations and a galaxy-halo connection model. Mock galaxies have several properties assigned from real data, from which we construct mock 
 data for our distance indicators: Fundamental plane and Tully-Fisher.}
Our mock samples represent the largest and most accurate sample of mocks used 
in cosmological studies of peculiar velocities to date. 


The paper is structured as follows. Section \ref{sec:desi} describes DESI and its peculiar velocity survey. Section~\ref{sec:bgs_base_mocks} presents the parent simulations used to build all mocks samples. The subsets of the parent mock galaxies used to define the density field used in clustering measurements is described in section~\ref{sec:bgs_clustering_mocks}. Sections~\ref{sec:fp_mocks} and \ref{sec:tf_mocks} describe the procedure to create realistic Fundamental plane and Tully-Fisher samples, respectively.
Section~\ref{sec:sn_mocks} explains how we assigned type-Ia supernovae to BGS galaxies. 
 {The production of catalogs suited for clustering measurements is explained in section~\ref{sec:clustering catalogs}} and lastly, in section~\ref{sec:consensus}, we introduce how we obtained a consensus result on the growth rate of structures 
from the three different methodologies \citep{DESIPV_Lai, DESIPV_Qin, DESIPV_Turner}.

\section{The Dark Energy Spectroscopic Instrument}
\label{sec:desi}

    At the Kitt Peak National Observatory in Arizona, USA, 
    the Dark Energy Spectroscopic Instrument (DESI) is a multiplex fiber-fed focal 
    plane mounted on the 4-meter Mayall Telescope \citep{desicollaborationDESIExperimentPart2016,desicollaborationDESIExperimentPart2016a,desicollaborationOverviewInstrumentationDark2022}.\footnote{DESI Collaboration Key Paper.}
    The DESI focal plane is equipped with a large corrector lens system \citep{millerOpticalCorrectorDark2024} 
    and 5,000 optical fibers which can be 
    automatically reconfigured with robotic positioners \citep{poppettOverviewFiberSystem2024}. 
    The spectrographs span the near-ultraviolet to near-infrared wavelengths 
    (3,600 to 9,800 \AA), delivering a spectral resolution that ranges 
    from 2,000 in the blue camera to 5,000 in the red camera. 
    The spectroscopic data reduction pipeline is described in \citet{guySpectroscopicDataProcessing2023} while the 
    observing strategy and operations are detailed in \citet{schlaflySurveyOperationsDark2023}. 
    DESI is now\footnote{A recent program redefinition will aim at observing beyond
    the original 5-year program over 14k deg$^2$.} a 8-year program aiming at observing 17k deg$^2$ of the Northern sky and measure nearly 63 million redshifts. 
    
    After validating its scientific program with early data  \citep{desicollaborationEarlyDataRelease2024,
    desicollaborationValidationScientificProgram2024}, 
    DESI has publicly released its Data Release 1 (DR1, \citealt{desicollaborationDataRelease12025}) in April 2025. 
    A series of cosmological results were also published with DR1, 
    including measurements of the baryon acoustic oscillations (BAO)
    and full-shape analysis of the galaxy two-point functions
    \citep{desicollaborationDESI2024VII2025,
    desicollaborationDESI2024VI2025, 
    desicollaborationDESI2024IV2025, 
    desicollaborationDESI2024II2025,
    desicollaborationDESI2024FullShape2025}.\footnote{DESI Collaboration Key Papers.} 
    While not yet public, Data Release 2 (DR2) was recently used to update 
    DESI BAO and cosmological constraints 
    \citep{desicollaborationDESIDR2Results2025,desicollaborationDESIDR2Results2025a}, 
    pointing towards an evolving dark energy model when in combination 
    with the cosmic microwave background or type-Ia supernova data. 
    This work will use the peculiar velocity survey of DR1. 

    \subsection{The DESI Bright Galaxy Survey}
    \label{sec:desi:bgs}
    
    Amongst the different tracers observed by DESI, we only employ 
    the $z<0.1$ range of the Bright Galaxy Survey 
    (BGS, \citealt{hahnDESIBrightGalaxy2022}). 
    The BGS is a flux-limited $r<19.5$ sample, 
    covering more than 10 million redshifts over $0< z < 0.6$. 
     {We employed a luminosity-limited subsample of the initial set of BGS galaxies at $z<0.1$ as point tracers of the density field, then used in cross-correlation with 
    peculiar velocity tracers.  }

    In sections~\ref{sec:bgs_base_mocks} and \ref{sec:bgs_clustering_mocks}
    we describe the production of synthetic versions of the DESI BGS 
    sub-sample, as well as the ``clustering'' versions, which include
    correct descriptions of the window function through random catalogs, 
    optimal weights and Malmquist bias corrections. 

    \subsection{The DESI Peculiar Velocity survey}
    \label{sec:desi:pv}

    The DESI Peculiar Velocity survey 
    is a secondary program aiming at producing the largest sample 
    of extragalactic distance measurements over $z<0.1$. 
    At the end of observations, we expect to obtain 
    133k distances to early-type elliptical galaxies
    using the Fundamental plane relation \citep{Dressler1987,djorgovskiFundamentalPropertiesElliptical1987} 
    and 53k distances to late-type spiral galaxies 
    using the Tully-Fisher relation \citep{tullyNewMethodDetermining1977}. 
    The target selection process of the DESI PV survey is described 
    in \citet{saulderTargetSelectionDESI2023}. 
    Fundamental plane distances require spectra with relatively high signal-to-noise ratio in order to obtain the central velocity dispersion of stars. 
    Tully-Fisher distances require asymptotic rotational velocities, so 
    in addition to a central fiber yielding the systemic redshift of the galaxy, 
    a second fiber is placed at $0.4R_{26}$ from its center, where $R_{26}$
    is the semimajor axis radius measured at the $\mu = 26$~mag arcsec$^{-2}$ 
    r-band isophote as provided in the Siena Galaxy Atlas (SGA). 
    Peculiar velocities are derived from the offsets to the average relations.
    
    Using DESI Early Data Release, the first samples of 4,191 FP and 1,136 TF distances 
    were analyzed  
    \citep{saidDESIPeculiarVelocity2024, douglassDESIEDRCalibrating2025}. 
    In DR1, we obtain peculiar velocities for 
    96,758 FP galaxies \citep{DESIPV_Ross} and 
    10,262 TF ones \citep{DESIPV_Douglass}, after several quality cuts. 
    The zero-point calibration of distances using type-Ia supernovae, 
    as well as an estimate of $H_0$ using DR1 distances, is 
    presented in \citet{DESIPV_Carr}. 
    
    In sections~\ref{sec:fp_mocks} and \ref{sec:tf_mocks} we describe
    the framework to produce mock catalogs for FP and TF samples, respectively.

    
\section{The DESI BGS parent sample mocks}
\label{sec:bgs_base_mocks}

With the goal of analyzing jointly the density and velocity fields to measure the 
growth rate of structures, it was essential to create self-consistent mock 
catalogs realistically reproducing both fields as observed by DESI. 
The starting point is the creation of a large sample of DESI BGS galaxies prior to any selection. 
We refer to those as the DESI BGS parent sample of mock galaxies, for which we closely followed the methodology from \cite{smithGeneratingMockGalaxy2024}. 
From the parent mocks, we built mocks for the DESI BGS clustering sample tracing the density field (section~\ref{sec:bgs_clustering_mocks}) and mocks for the DESI FP and TF samples tracing the velocity field (sections~\ref{sec:fp_mocks} and \ref{sec:tf_mocks}). 

    \subsection{The AbacusSummit suite of n-body simulations}
    \label{sec:bgs_base_mocks:abacus}
    
    We build the parent mock catalogs from the \textsc{AbacusSummit} suite of n-body simulations \citep{garrisonABACUSCosmologicalNbody2021, maksimovaABACUSSUMMITMassiveSet2021}. 
    Those pure matter simulations evolve $6912^3$ particles with mass of $2 \times 10^9 h^{-1} M_\odot$ in a box of 2$h^{-1}$Gpc, assuming the Planck 2018 best-fit flat $\Lambda$CDM cosmology \citep{planckcollaborationPlanck2018Results2020}: 
    $h = 0.6736$, 
    $\omega_\mathrm{cdm} = 0.1200$, 
    $\omega_b = 0.02237$,
    $\sigma_8 = 0.8114$, and $n_s = 0.9649$.
    There are a total of 25 realizations of different initial conditions, drawn from 
    second-order Lagrangian perturbation theory at $z=99$. 
    Another 96 cosmological models have  also  been simulated, using the same initial conditions, 
    though those were not used in this work. 
    \textsc{AbacusSummit} is the largest high-accuracy cosmological n-body data set produced to date. 
    All data products are publicly available.\footnote{\url{https://abacussummit.readthedocs.io/}}

     {A catalog of halos is provided with the simulation. Halos were identified} using an improved implementation of the friends-of-friends and the spherical overdensity algorithms, named \textsc{CompaSO} \citep{hadzhiyskaCOMPASONewHalo2022}. 
    To avoid non-physical unresolved halos, 
    a mass cut of $M_h > 10^{11} h^{-1} M_\odot$ was applied. 
     {However, halos of lower masses are required to produce the full parent BGS sample.  
    Following \citet{smithGeneratingMockGalaxy2024}, 
    we solve this resolution issue by
    randomly selecting field
    particles (not assigned to halos) to be used as the locations of 
    halos below this mass cut. }
    Halo masses are assigned to field particles by 
    extrapolating the halo mass function to lower masses, 
    assuming a power-law. 
     {The velocity of the field particles are directly assigned to their 
    respective halo. }
    \citet{smithGeneratingMockGalaxy2024} show that 
    this is a good approximation based on clustering statistics of 
    field particles versus halos. 

    \subsection{Galaxy assignment scheme}
    \label{sec:bgs_base_mocks:hod}
    
    We populate halos with galaxies using a luminosity-dependent halo occupation distribution (HOD) model \citep{smithLightconeCatalogueMillenniumXXL2017,smithLightconeCatalogueMillenniumXXL2022}, reproducing the comoving number density of galaxies as well as the projected correlation function of the DESI DR1 BGS sample. 
    The best-fit parameters of this ``nested-HOD'' model were previously determined using the $z=0.2$ snapshot of the \textsc{AbacusSummit} halos and the real DESI BGS galaxies over $0.1 < z < 0.6$. 
     {This snapshot redshift and the real data redshift range} were used to build the official BGS mock catalogs used in the DESI BAO and full-shape analyses. 
     {Since this procedure was extensively tested,} here we simply assume that this model can be extrapolated to the 
    redshifts of our interest, i.e. $z<0.1$.\footnote{
    Ideally one should be able to refit the HOD parameters using the 
    $z=0.1$ \textsc{AbacusSummit} snapshot (the lowest redshift available) 
    and BGS galaxies over $0<z <0.1$, but we decided to stick 
    to the already available and tested HOD model fitted over $0.1 < z < 0.6$ using the $z=0.2$ snapshot. We leave this improvement to future work. 
    }
    When performing cosmological inference, we carefully
    compare cosmological results to their predictions computed at $z=0.2$, 
    which corresponds to the redshift of the underlying matter clustering in our mock catalogs.
     {Additionally, given the small change in the value of $\fsig$ between $z=0.2$ and $0.1$, we expect the impact on clustering to be insignificant given our uncertainties.} 

    We placed central galaxies at the center of their halo, 
    and we assign them the same velocity. 
    The number of satellite galaxies above a minimum luminosity
    threshold is drawn from a Poisson distribution. 
    We position satellite galaxies following a NFW profile, 
    with a random virial velocity drawn 
    from a Gaussian with velocity dispersion $\sigma^2 = G M_{200}/2R_{200}$
    (in each direction),  {where $M_{200}$ and $R_{200}$ are the mass and radius
    corresponding to an overdensity of $\delta=200$. }
    We assume that such dispersion is constant in this work. 

    Additionally to positions and velocities, we draw absolute magnitudes in the $r$-band $M_r$
    and rest-frame colors $(g-r)$ from a realistic redshift-evolving luminosity function \citep{smithLightconeCatalogueMillenniumXXL2017,smithLightconeCatalogueMillenniumXXL2022}. 

    \subsection{Cut-sky mock catalogs}
    \label{sec:bgs_base_mocks:cut_sky}
    
    After placing the observer in the box, 
    we convert relative positions and velocities to sky coordinates 
    (right-ascension, declination) and redshift. 
    These are often referred to as ``cut-sky'' mocks. 
    Both cosmological and peculiar components of the redshift are provided. 
    The first one, $z_\mathrm{cos}$, is computed from the assumed cosmology, 
    while the second one, $z_\mathrm{v} = (\vec{v}\cdot \hat n)/c$ is 
    computed from the radial component of the 3D peculiar velocity 
    $\vec{v}$ with 
    respect to the line-of-sight of the observer $\hat{n}$. 
    Both redshift components are combined into the observed
    redshift following
    \begin{equation}
        1 + z_\mathrm{obs} = (1+z_\mathrm{cos}) (1+z_\mathrm{v}).
        \label{eq:obs_redshift}
    \end{equation}
    The absolute magnitudes and rest-frame colors are then converted 
    to apparent magnitudes and observed-frame colors using the 
    cosmological redshift.

    Since we are interested in the very low-redshift Universe ($z<0.1$), 
    we divide the initial 2 $h^{-1}$~Gpc simulation boxes into 
    $3^3$ sub-volumes, placing the observer at the center of each.
    This allows us to produce a total of $25 \times 27 = 675$ mock 
    realizations. This is the largest sample of peculiar velocity mocks
    based on n-body simulations to date.
    Since they are cut from the same initial box, all 27 sub-volumes 
    are not exactly independent, particularly concerning peculiar velocities
    which are are correlated on large scales.
    Appendix A of \citet{carreresGrowthrateMeasurementTypeIa2023}, who 
    uses similar methodology to produce mocks, 
    estimates this correlation between neighboring sub-volumes 
    to be below 1.6\%, 
    so they can be safely neglected. 
    The resulting cut-sky mocks 
    cover the full sky, to which we apply observational completeness 
    masks corresponding to DR1 observations \citep{desicollaborationDESI2024II2025}. 

\begin{figure*}
\centering
\includegraphics[width=0.9\textwidth]{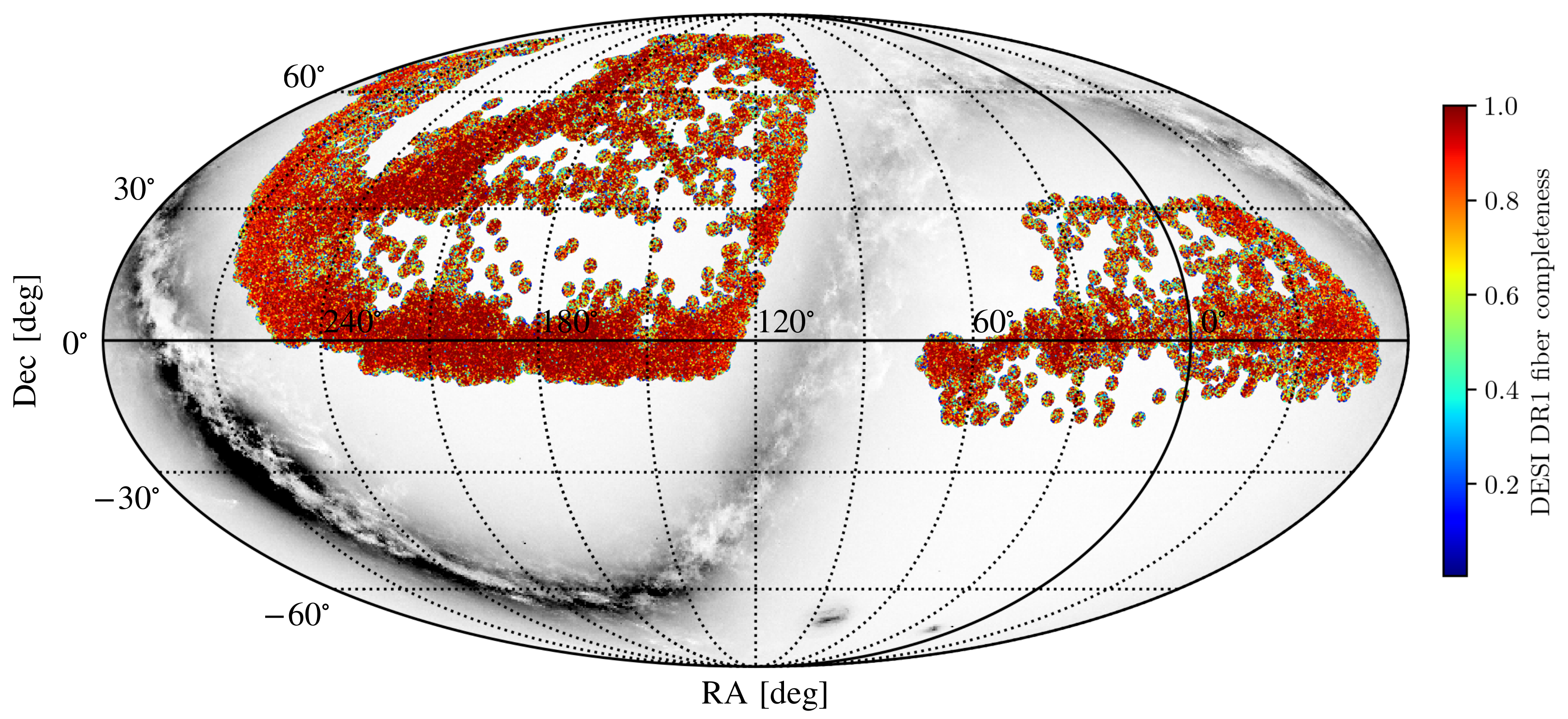}
\caption{DESI DR1 sky coverage of mock BGS galaxies with color indicating spectroscopic completeness. }
\label{fig:footprint}
\end{figure*}

    Figure~\ref{fig:footprint} shows the sky distribution of a mock BGS realization after considering regions where the observational completeness is non-zero. 

    \subsection{Additional galaxy properties}
    \label{sec:bgs_base_mocks:properties}
    
    We add ancillary properties such as star formation rates, metallicities and velocity dispersions to our parent BGS mock galaxies 
    through an additional nearest-neighbor matching to the real 
    DESI BGS data. 
    We build a $k$-d tree to assign values for
    \texttt{TARGETID}, \texttt{SURVEY}, \texttt{PROGRAM} and \texttt{HEALPIX} from real galaxies using the nearest mock galaxies in the space of $r$-band absolute magnitude, $(g-r)$ color and redshift.
    This combination of identifiers allows us to recover additional properties
    from real data while preserving correlations between such quantities. 
    
\begin{figure*}[t]
        \centering
        \includegraphics[width=0.49\linewidth]{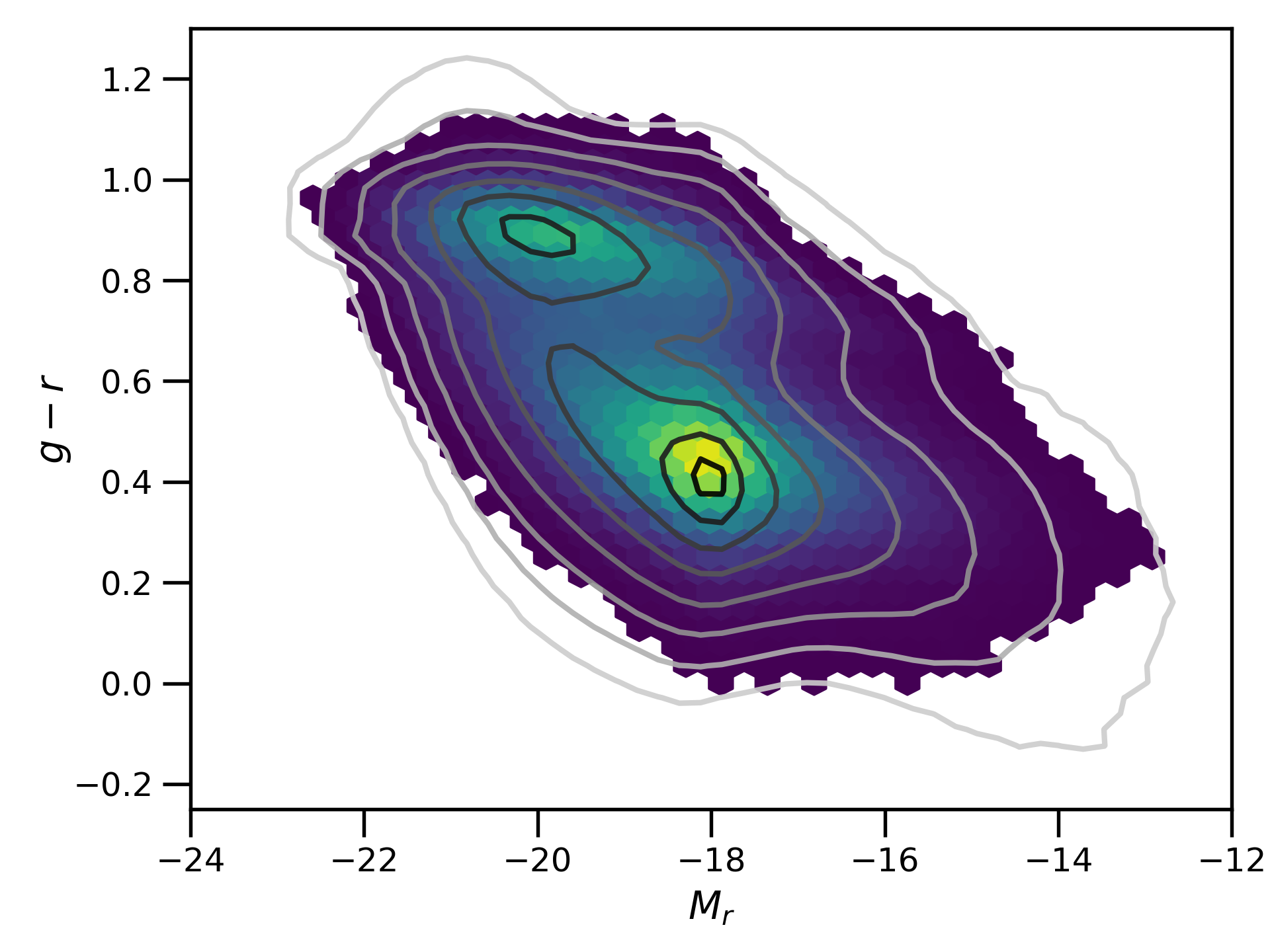}
        \includegraphics[width=0.49\linewidth]{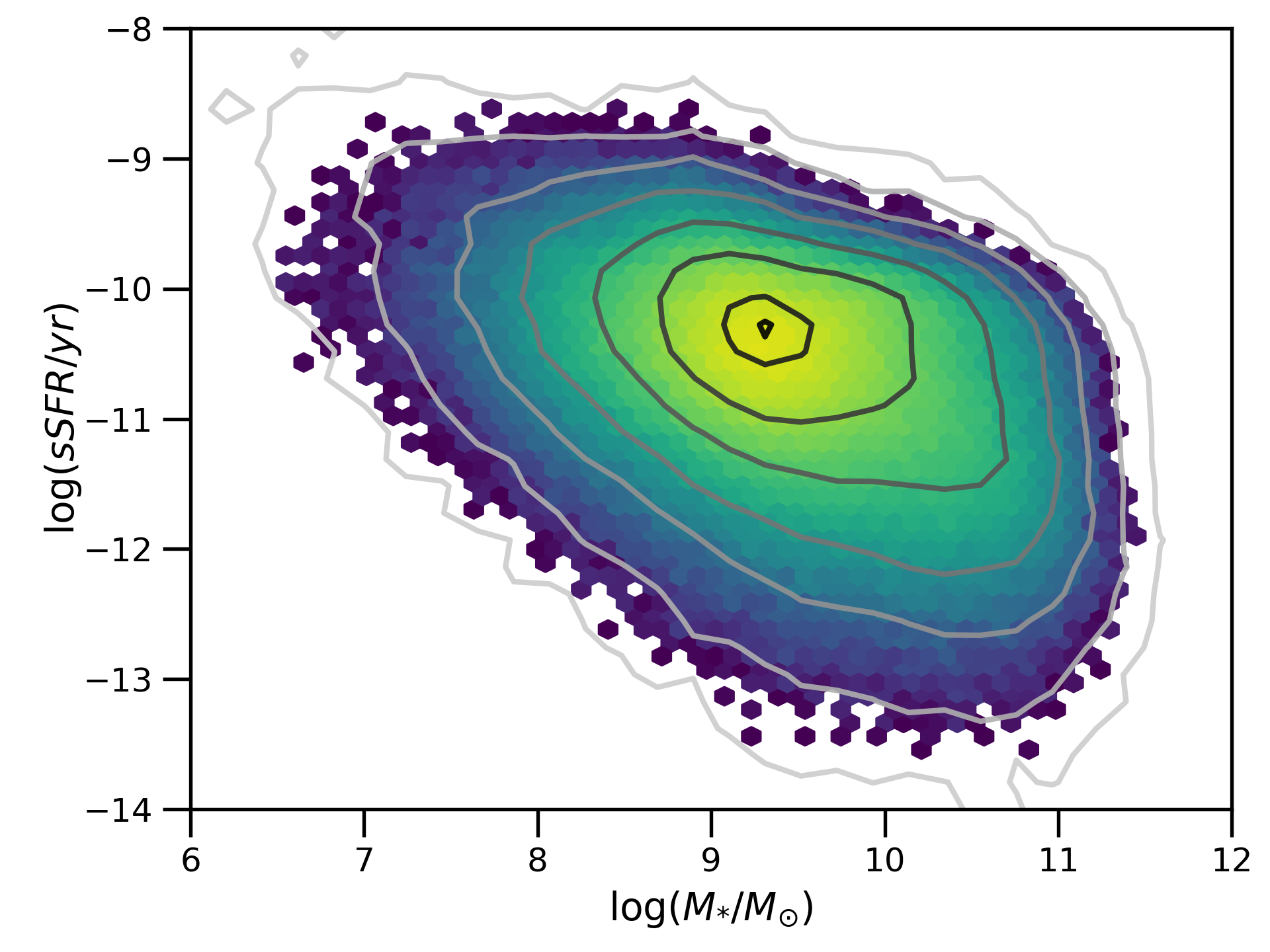}
        \caption{Example comparisons between DESI BGS data and the base AbacusSummit mock catalogs. The left panel shows the $r$-band absolute magnitude against $g-r$ color, while the right panel shows the stellar mass against specific star formation rate. In both cases, the contours show the average properties over 27 simulations, with dark contours indicating regions of higher density. The colormap provides the number of real DESI galaxies in each bin, with light colours indicating higher density. For the purposes of this comparison, we restrict the DESI data to only those galaxies with $z<0.11$ and apparent magnitude $r<19.7$, consistent with the simulations.}
        \label{fig:bgs_iron_mocks}
    \end{figure*}

    Figure~\ref{fig:bgs_iron_mocks} shows an example of the color-magnitude 
    diagram and the stellar mass versus specific star formation rate for the DESI BGS sample (cut to the same redshift and apparent magnitude limits as our mocks) against the mocks. 
    There is excellent agreement across both of these relationships, 
    both in terms of the range of the parameter space spanned by both the data and mocks, and in the relative densities with which the real and simulated galaxies populated these spaces. 

    \subsection{Random catalogs}
    \label{sec:bgs_base_mocks:randoms}

    The window function of the survey is described with a uniform sampling of random positions in the volume. 
    We first draw uniformly distributed sky coordinates over the sphere, 
    then we assign to each random point all other quantities (redshift, magnitudes, colors) from a randomly selected mock galaxy. 
    This ensures that the distribution of random redshifts (and other quantities) follows exactly the one from the clustered mock galaxies.
    
    For each survey realization, we draw as many random points as
    there are galaxies in the mock. For clustering measurements, 
    several random catalogs are combined into a single catalog 
    to reduce shot noise. 

    Finally, we assign fiber completeness values from the real data
    to the randoms according to their sky position. Typically 
    fiber completeness corrections are performed as weights applied 
    to the randoms.

\section{The BGS clustering mocks}
\label{sec:bgs_clustering_mocks}

    This section describes the production of the mock galaxy catalogs used as probes of the matter overdensity field, which we sub-sample from the parent BGS mock catalogs (section~\ref{sec:bgs_base_mocks}).  
    We restrict the mocks to the same redshift range $0.01 < z < 0.10$ as the low-redshift BGS dataset analyzed in this study, apply an apparent magnitude cut $r < 19.5$ matching the BGS BRIGHT selection, 
    and randomly sub-sample the parent dataset as a function of sky position in accordance with the BGS angular completeness mask (Figure~\ref{fig:footprint}) corresponding to the DR1 sample \citep{desicollaborationDESI2024II2025}. 
    We apply these same selections to the random catalogs accompanying the mocks.
    
     {Similarly to the BAO and full-shape analysis of real data, we apply an absolute magnitude cut to the galaxy sample to obtain a number density of traces that is nearly constant over the considered redshift range.
    The chosen cut in the absolute magnitude of real galaxies in the $r$-band is $M_r < -17.7$.  } 
    The $M_r$ value is determined from the $k$-corrected SDSS $r$-band absolute magnitude, $M_{r,{\rm fsf}}$, found using {\sc Fastspecfit} \citep{moustakasFastSpecFitFastSpectral2023}, and an evolution correction based on the redshift $z$,
    \begin{equation}
        M_r = M_{r,{\rm fsf}} + 0.97 \, z - 0.095 ,
    \end{equation}
    following \cite{desicollaborationDESI2024II2025}.  
    Whereas the BGS sample used in the DESI Baryon Acoustic Oscillation analysis is selected by $M_r < -21.5$, we apply a fainter luminosity cut $M_r < -17.7$, given the lower redshift of our galaxy selection and different signal-to-noise characteristics of our science goal.
    
\begin{figure}[t]
        \centering
        \includegraphics[width=\linewidth]{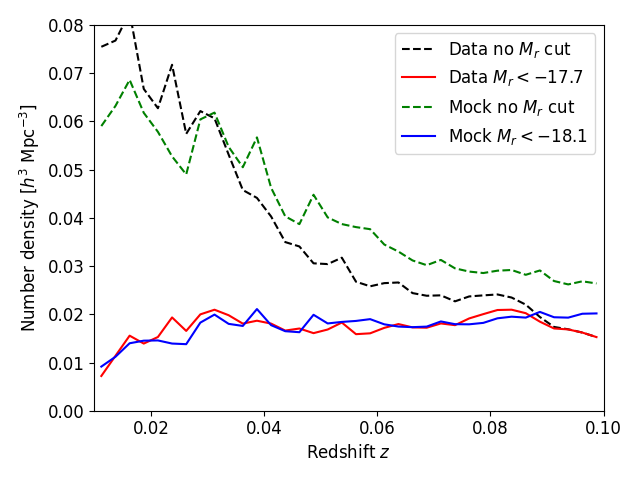}
        \caption{The number density versus redshift of DESI BGS galaxies in the density-field sample compared to a sample mock catalog, before and after applying a luminosity cut.}
        \label{fig:density_nz}
    \end{figure}

     {Figure~\ref{fig:density_nz} displays the number density of the density-field DR1 data and mocks, before and after the application of the luminosity cut. 
    We see small differences in the number density of parent tracers (dashed lines) which are due to the extrapolation of the HOD models to our lower redshift range.
    Therefore, we applied a slightly different luminosity cut to mocks, tuned as $M_r < -18.1$. This choice results in matching number densities for mocks and data density fields. 
    }
    We further sub-sampled the random catalogs to match the redshift-dependent luminosity cut.
    
    The determination of optimal weights for clustering measurements requires the number density of the galaxy density field at each location.  We determined the average number density of the mock dataset on a $128^3$ grid by combining together all 675 mocks, and applied this gridded number density to each object in the mock galaxy and random catalogs. For the real data, we construct the selection function by stacking and gridding the random catalogs.
    
\section{The Fundamental Plane mocks}
\label{sec:fp_mocks}
    
    This section describes how the parent BGS mock galaxy catalogs (section~\ref{sec:bgs_base_mocks}) are sub-sampled to produce mocks representative of the DESI DR1 Fundamental Plane (FP) sample described in detail in \citet{DESIPV_Ross}. 
    These sub-samples are essential for validating and characterizing the uncertainties of both the FP fits in that work, and subsequent clustering measurements using FP peculiar velocities. 
    
    The Fundamental Plane is a relationship between the effective radius, $r$, surface brightness, $i$, and velocity dispersion, $s$, of elliptical galaxies. The FP mocks hence need to reproduce the distributions of these parameters seen in the real data. Given the FP data is used primarily as a tracer of the velocity field, it is also important that the mocks reproduce the selection function, number density, log-distance ratio and log-distance ratio error distributions seen in the real data. This section demonstrates that this is the case.
    
    \subsection{Basic selection}
    \label{sec:fp_mocks:data_cuts}

        The mock generation procedure used in this work is loosely based on the one used to produce mocks for the Sloan
        Digital Sky Survey FP sample (\citealt{howlettSloanDigitalSky2022}, H22 hereafter) 
        and references therein. A significant difference
        is that H22 fits a HOD-style model and populates 
        simulated halos with FP galaxies directly. 
        Such an approach is not feasible here due 
        to the requirement of having to produce both FP and TF subsamples on top of a larger redshift sample. 
        
        The FP subsamples in this work start from ancillary properties for each BGS mock galaxy derived from the cross-match to the real data as described in Section~\ref{sec:bgs_base_mocks:properties}. 
        We then apply the following additional selection cuts:
        \begin{enumerate}[itemsep=4pt]
            \item{Heliocentric redshift $0.0033<z<0.1$;}
            \item{Morphological type either DeVaucoleurs or S\'ersic with S\'ersic index $n_{s}>2.5$;}
            \item{Axial ratio $b/a > 0.3$;}
            \item $r$-band apparent magnitude $10.0 < m_{r} < 18.0$;
            \item $k$-corrected colour cuts with $g-r > 0.68$ and $$1.3(r-z)-1.2<g-r<2.0(r-z)-0.15.$$
        \end{enumerate}
        These are almost identical to those used in the real FP data, with only small changes to the color cuts. 
        The limited halo mass resolution of the AbacusSummit simulations previously discussed in  {section~\ref{sec:bgs_base_mocks:abacus}} results in a lack of mock FP galaxies for $z<0.04$ when the original data color cuts are used.  {This is accounted for by using $k$-corrected color cuts for the mocks (where the original data cuts were applied using non $k$-corrected colors), and shifting the intercept of the upper color cut from $-0.05$ for the data to $-0.15$ for the mocks.}
        
\begin{figure}[t]
            \centering
            \includegraphics[width=\linewidth]{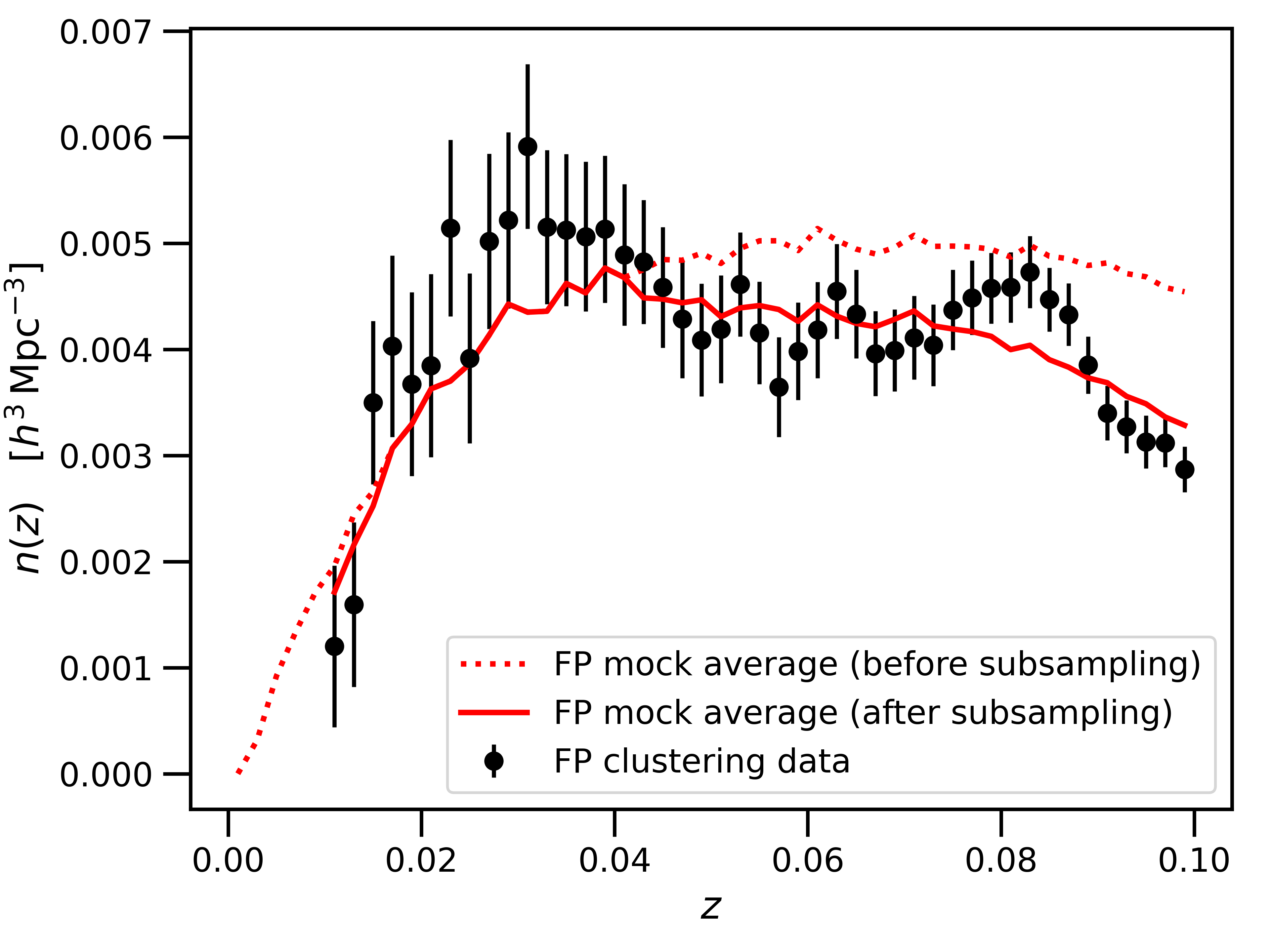}
            \caption{The number density versus redshift of Fundamental Plane galaxies in the data compared to 
            the average of 675 mocks, before (dotted) and after (solid line) a subsampling correction is applied to ensure the mocks match the expected number of galaxies. The error bars are derived from the standard deviation of over the 675 mocks and used to express the expected uncertainty on the data.}
            \label{fig:fp_nz}
        \end{figure}

        Following this change, the average number density as a function of redshift matches closely that of the data, albeit with a small excess at higher redshifts.  
        We randomly subsample mock galaxies based on the difference between the smoothed data and mock mean. 
        The number density both before and after subsampling is shown in Fig~\ref{fig:fp_nz}.
        
    \subsection{Generating mock measurements of FP parameters}
    \label{sec:fp_mocks:noisy_fp_parameters}

        The next step is to generate observed FP properties for each mock. 
        While the cross-match to the real data also provides effective radii $r$, surface brightnesses $i$ and velocity dispersions $s$, the same resolution issues as mentioned above (and the resulting changes to the color cuts) were found to result in mock FP samples that did not adequately match that found in the data. 
        Instead, we copied again the method of H22 and assigned values of $r$, $s$ and $i$ to each galaxy by drawing from a 3D Gaussian distribution. 
        However, unlike that previous work, this random sampling 
        is tied to the absolute magnitude of each galaxy. 
        
        More precisely, the $r$-band absolute magnitude of a galaxy can be related to the log effective radius $r$ and log $i$-band surface brightness $i$ via
        \begin{equation}
            M_{r} = M_{r,\odot} - 5r - 2.5i - 2.5\log{2\pi} - 15,
            \label{eq:FP_abs_mag}
        \end{equation}
        where $M_{r,\odot}=4.65$  {is the $r$-band absolute magnitude of the Sun.}
        In addition, the fundamental plane is usually approximated as a 3D Gaussian with mean $\mu = \{\bar{r},\bar{s},\bar{i}\}$ and covariance matrix 
        \begin{equation}
            \boldsymbol{\mathsf{C}}_{n} = \boldsymbol{\Sigma} + \boldsymbol{\mathsf{E}}_{n} = 
            \begin{pmatrix}
            \sigma^{2}_{r} & \sigma_{rs} & \sigma_{ri} \\
            \sigma_{rs} & \sigma^{2}_{s} & \sigma_{si} \\
            \sigma_{ri} & \sigma_{si} & \sigma^{2}_{i} 
            \end{pmatrix} + 
            \begin{pmatrix}
            e^{2}_{r}+\epsilon^{2}_{r} & 0 & -e_{r}e_{i} \\
            0 & e^{2}_{s} & 0 \\
            -e_{r}e_{i} & 0 & e^{2}_{i}
            \end{pmatrix} ~.
            \label{eq:covariance}
        \end{equation}
        which has contributions from both the intrinsic scatter of the FP ($\Sigma$) and observational error ($E_{n}$). $e_{r}$, $e_{s}$, $e_{i}$ denote the observational uncertainties on the $r$, $s$ and $i$ parameters and $\epsilon_{r} = \log_{10}(1+300/cz)$ is a small additional contribution to the uncertainty on the effective radius for each galaxy, based on its observed redshift $z$, to account for non-linear peculiar velocities. 
        For the purposes of generating true FP parameters for each mock galaxy we consider only the component $\Sigma$.
        
        \begin{figure*}[t]
            \centering
            \includegraphics[width=0.33\linewidth]{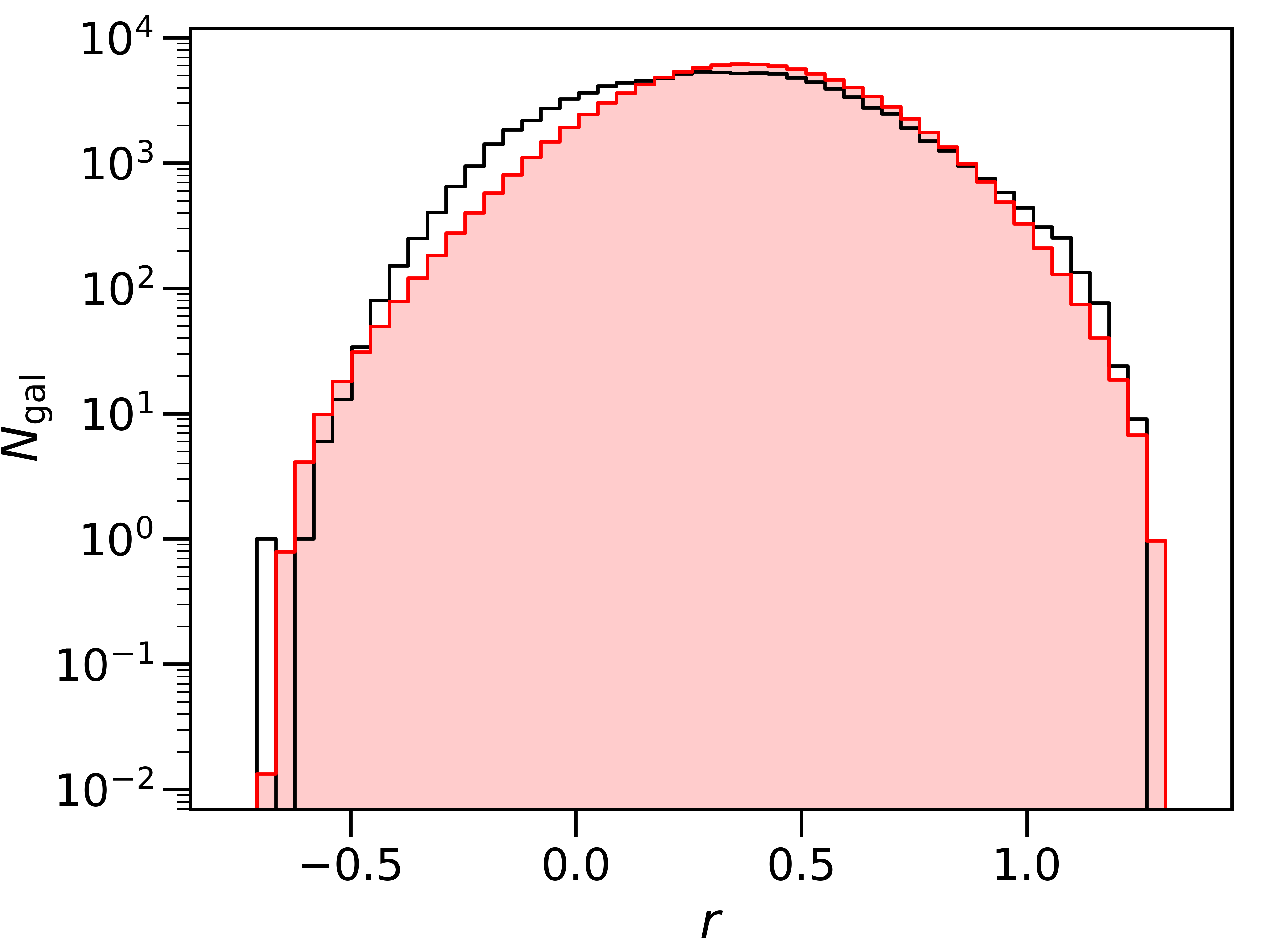}
            \includegraphics[width=0.33\linewidth]{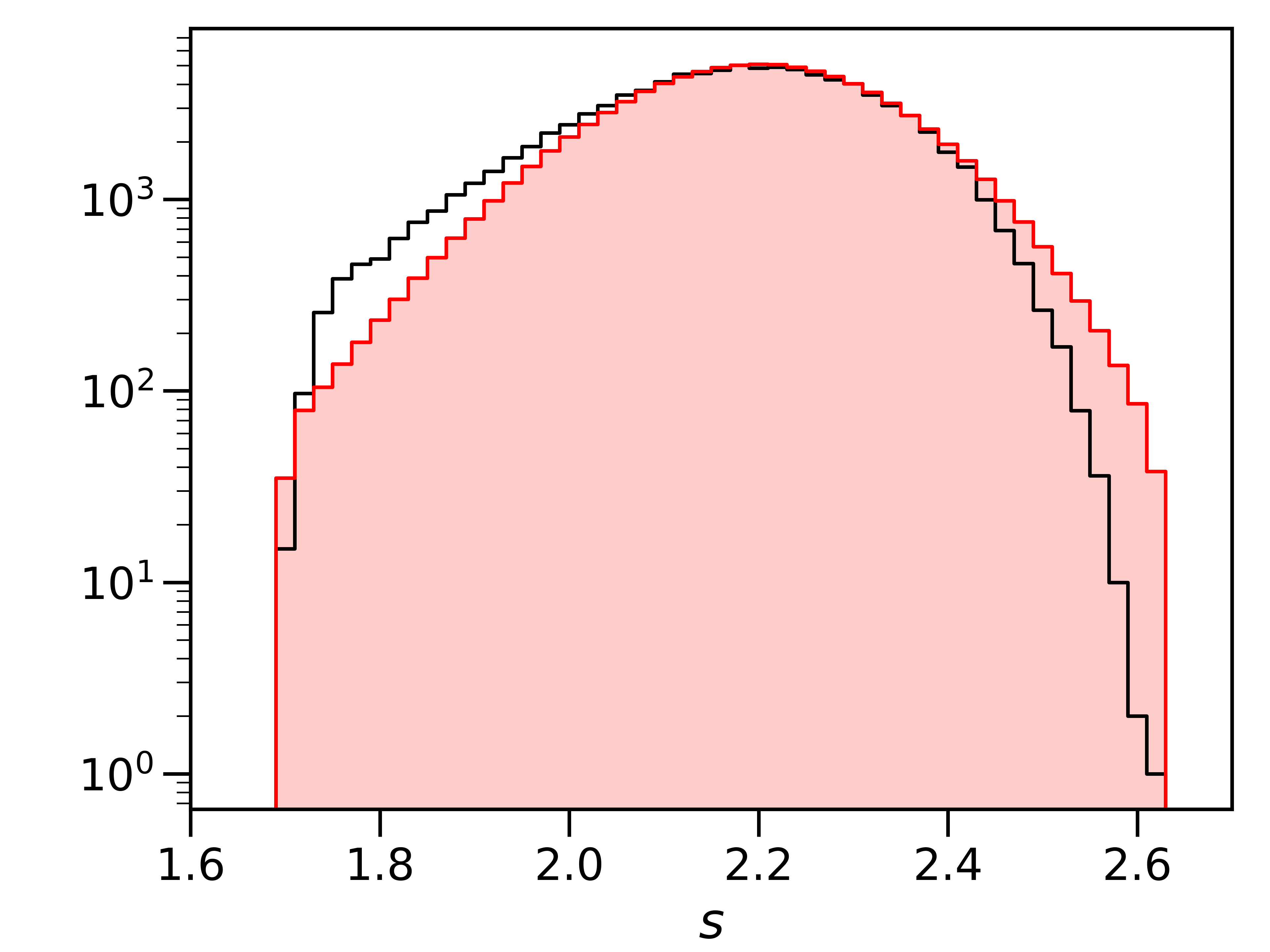}
            \includegraphics[width=0.33\linewidth]{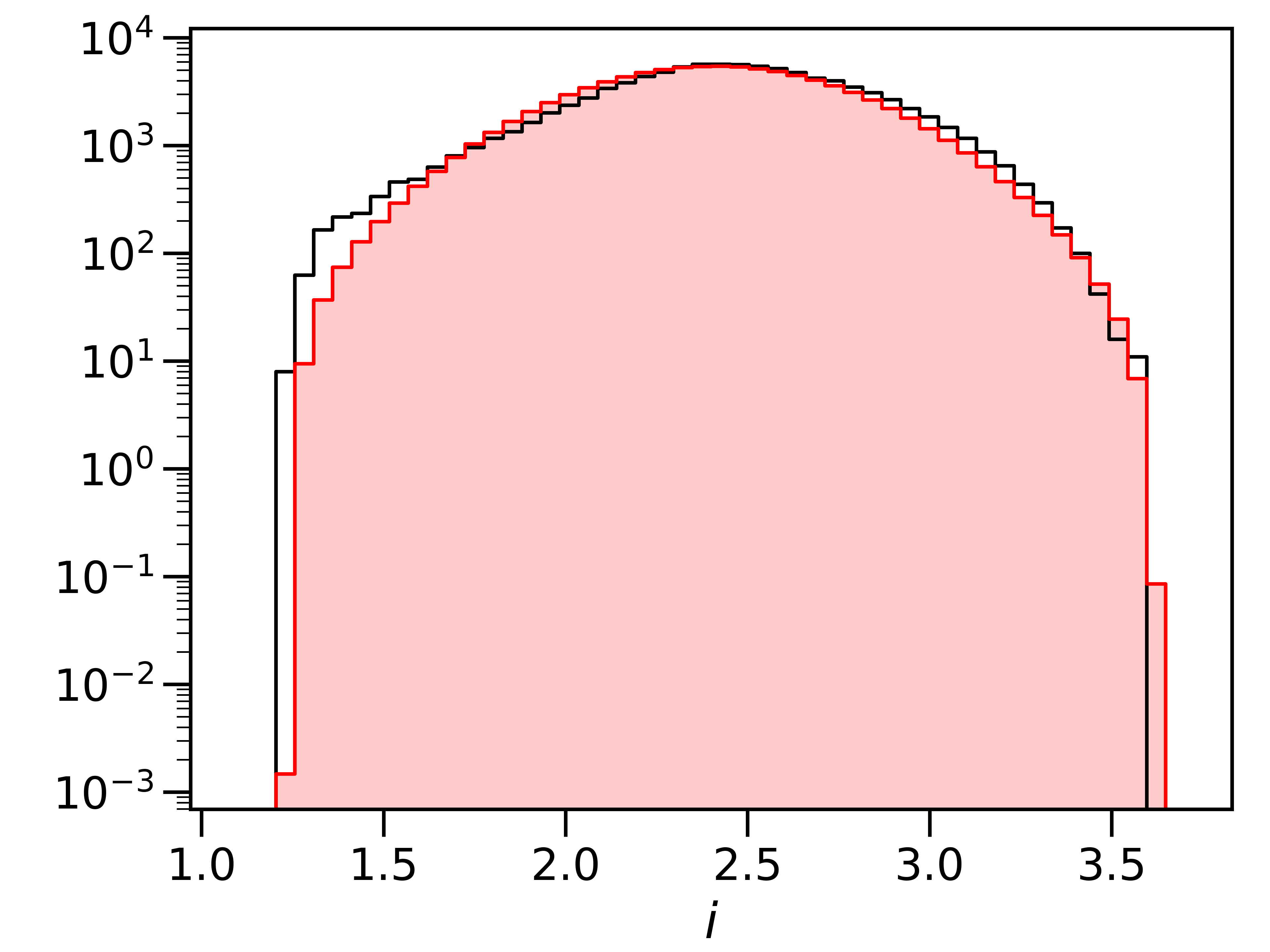} \\
            \includegraphics[width=0.33\linewidth]{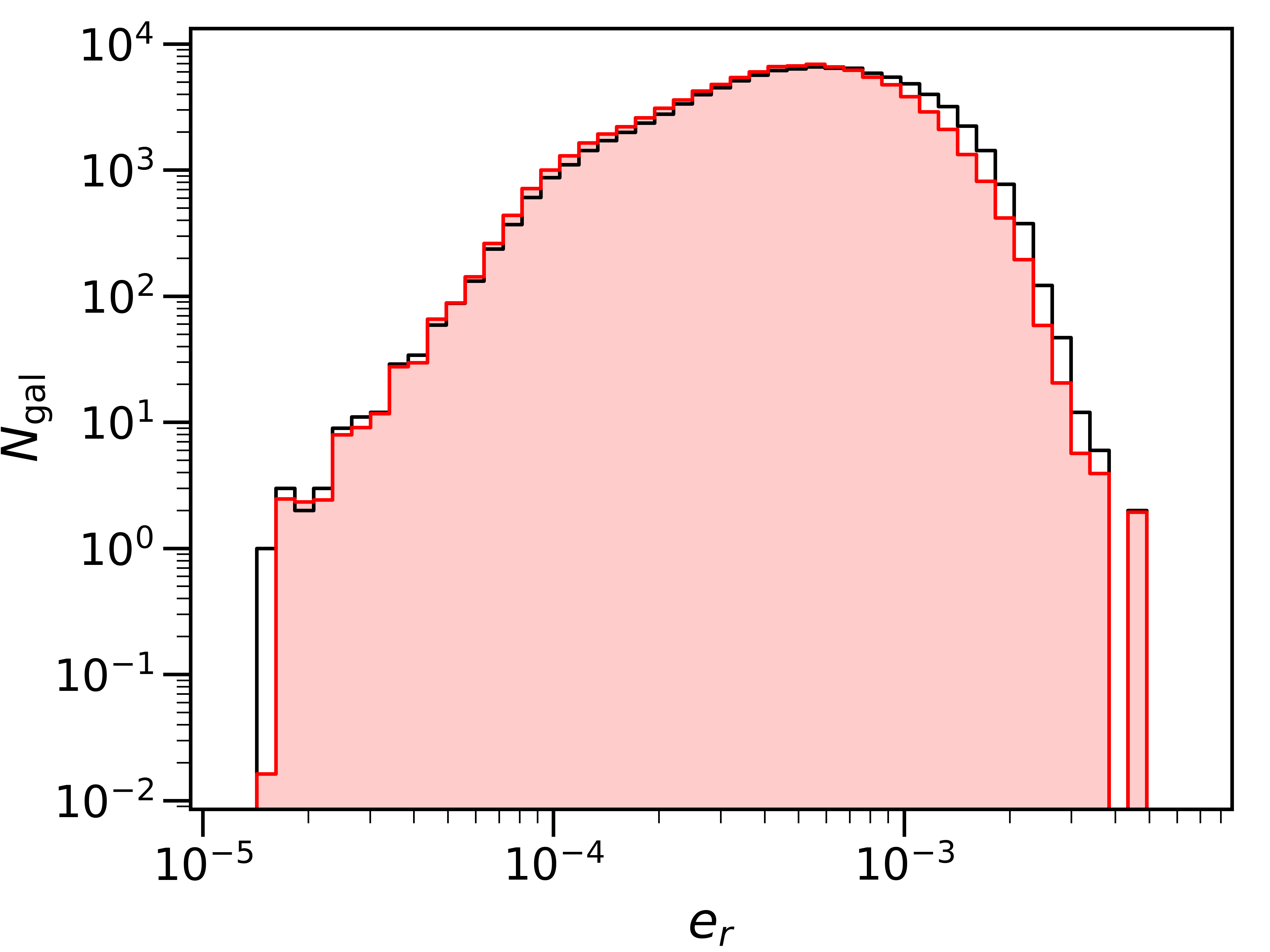}
            \includegraphics[width=0.33\linewidth]{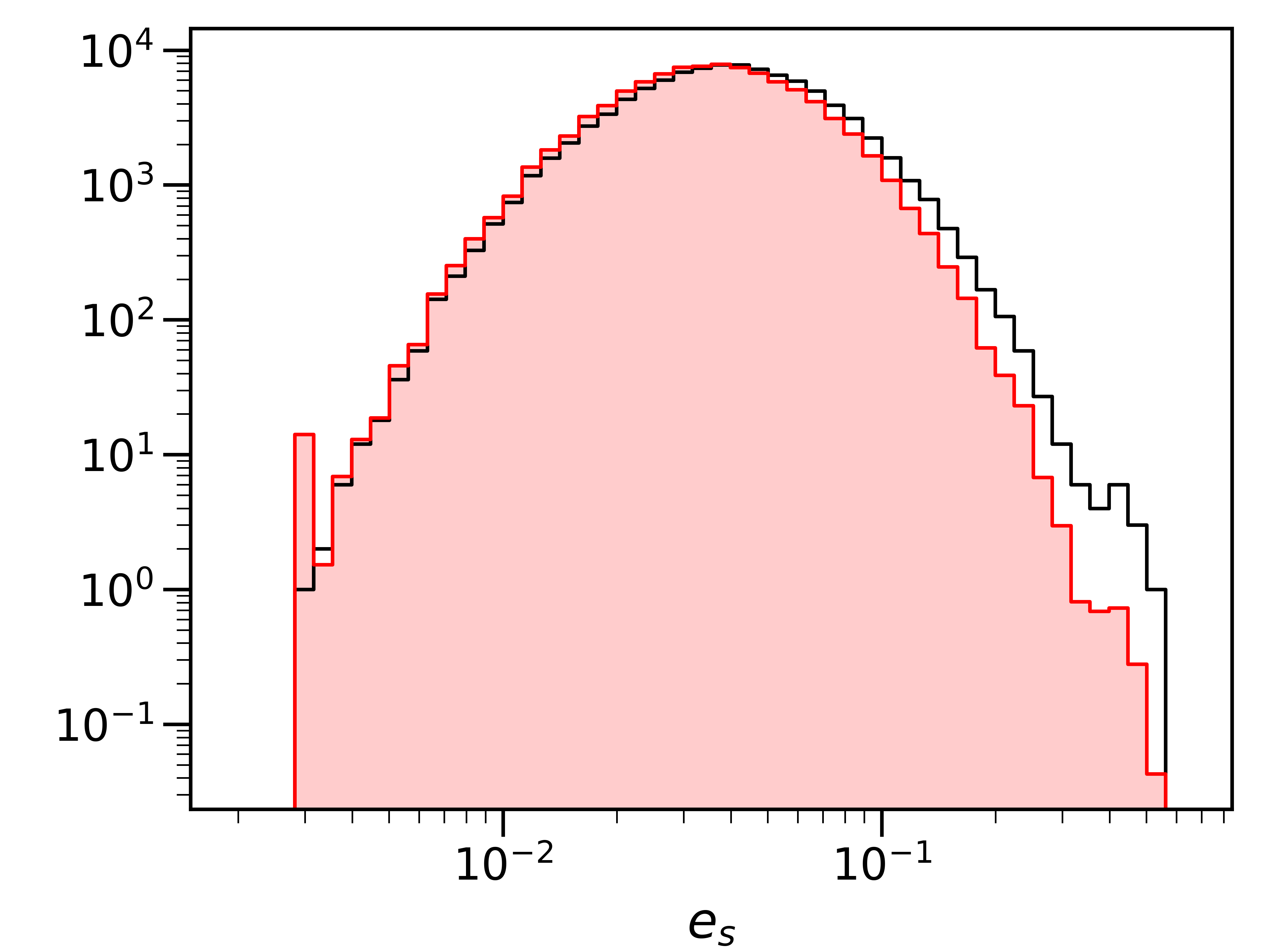}
            \includegraphics[width=0.33\linewidth]{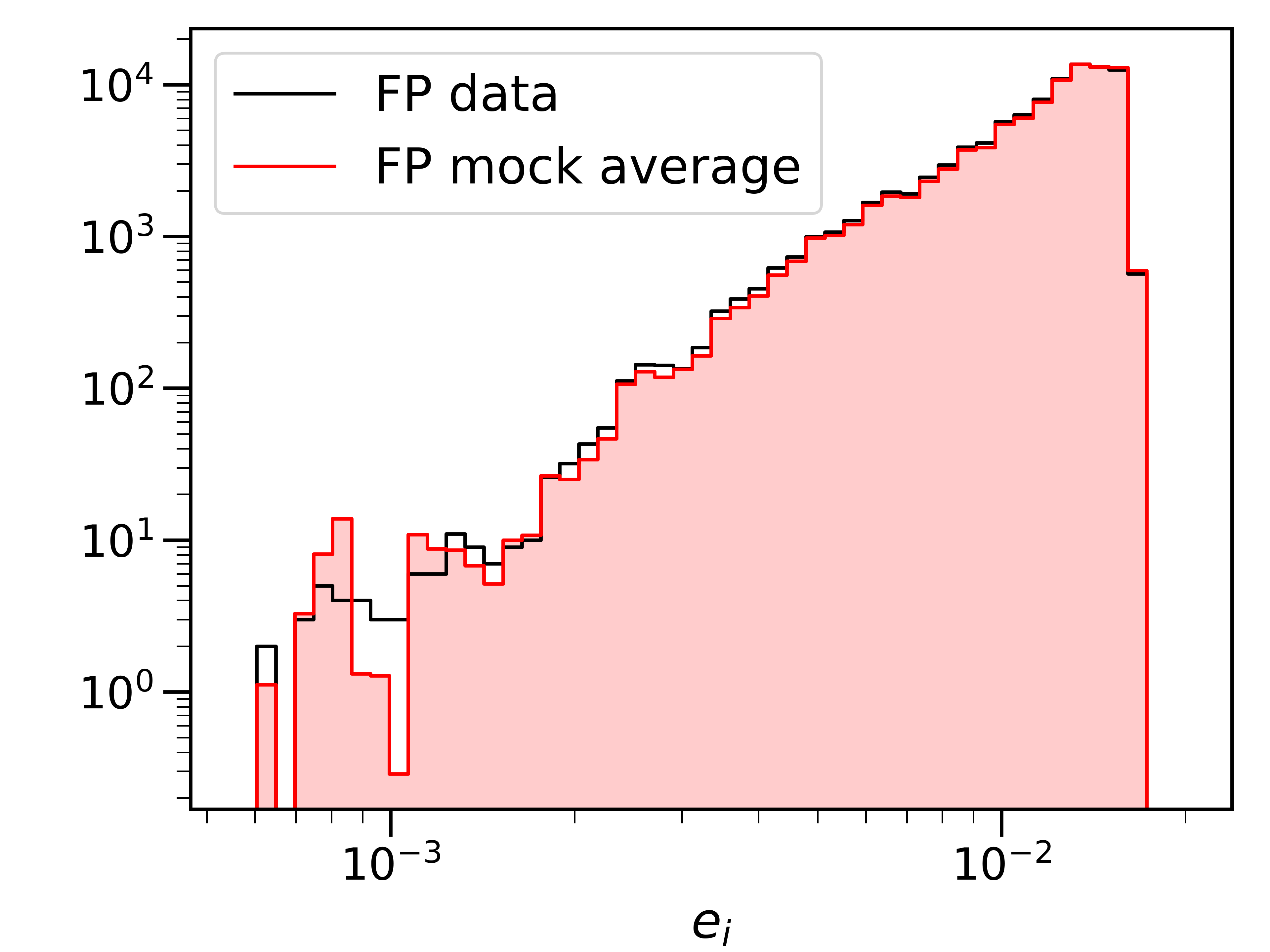}
            \caption{Distribution of Fundamental Plane input parameters (effective radius $r$, velocity dispersion $s$ and surface brightness $i$) and their uncertainties for both the DESI FP data and averaged over 625 FP mock catalogs.}
            \label{fig:fp_mock_inputs}
        \end{figure*}
        
        Given this, the fundamental plane can first be recast in terms of absolute magnitude, and then conditioned on this property. The result is a 2D Gaussian $P(s, i | M_{r})$ with mean $\mu=\{\hat{s}, \hat{i}\}$ and covariance matrix $\hat{\Sigma}$. The new mean can be written in terms of the recast FP parameters using
        \begin{equation}
            \hat{i} = \bar{i} + \frac{\sigma_{M_{r},i}}{\sigma^{2}_{M_{r}}}(M_{r}-\bar{M}_{r}); \qquad
            \hat{s} = \bar{s} + \frac{\sigma_{M_{r},s}}{\sigma^{2}_{M_{r}}}(M_{r}-\bar{M}_{r}),
        \end{equation}
        where the mean absolute magnitude is evaluated from the mean surface brightness and velocity dispersion via Eq.~\ref{eq:FP_abs_mag} and 
        \begin{align}
           \sigma^{2}_{M_{r}} &= 25\sigma^{2}_{r} + 25\sigma_{r,i} + 6.25\sigma^{2}_{i} \\
           \sigma_{M_{r}i} &= -5\sigma_{r,i} -2.5\sigma^{2}_{i} \qquad  \sigma_{M_{r}s} = -5\sigma_{r,s} -2.5\sigma_{s,i}. 
           \label{eq:covariance_trans}
        \end{align}
        The covariance matrix for the conditional probability is
        \begin{equation}
            \boldsymbol{\hat{\Sigma}} = 
            \begin{pmatrix}
            \sigma^{2}_{s} - \frac{\sigma^{2}_{M_{r}s}}{\sigma^{2}_{M_{r}}} & \sigma_{si} - \frac{\sigma_{M_{r}s}\sigma_{M_{r}i}}{\sigma^{2}_{M_{r}}} \\
            \sigma_{si} - \frac{\sigma_{M_{r}s}\sigma_{M_{r}i}}{\sigma^{2}_{M_{r}}} & \sigma^{2}_{i} - \frac{\sigma^{2}_{M_{r}i}}{\sigma^{2}_{M_{r}}} \\
            \end{pmatrix}.
        \end{equation}

        Using the above PDF, `true' values for $s$ and $i$ are drawn for each mock galaxy, and then used alongside Eq~\ref{eq:FP_abs_mag} to produce $r$. For the sampling, we fix the fundamental plane parameters $a$, $b$, $\bar{r}$, $\bar{s}$, $\bar{i}$, $\sigma_{1}$, $\sigma_{2}$ and $\sigma_{3}$ (which control the mean, orientation and intrinsic scatter of the FP) to the fiducial best-fit values of real data \citep{DESIPV_Ross}. 
        
        The next stage is to incorporate the effect of the true peculiar velocity/log-distance ratio on each galaxy's observed size using the transformation $r \rightarrow r + \eta$, and then assign observational uncertainties and produce observed quantities. This is done by again using nearest neighbor matching --- for each mock galaxy we find the nearest real FP galaxy in the 3D $\{r, s, i\}$-space and assign its uncertainty. 
        This ensures that correlations between the measurements and the errors in the real data are also represented in the mocks. 
        Observed values are generated by drawing from a 3D Gaussian for each galaxy with mean equal to the true $r$, $s$, and $i$ values, and covariance given by $E_{n}$ in Eq~\ref{eq:covariance}. 
        Finally, one last selection cut is applied to the mocks of $\log_{10}(50) < s < \log_{10}(420)$, which again matches that applied to the data to ensure only robust velocity dispersion measurements are used.
        
        The outcome of this procedure is summarized in Figure~\ref{fig:fp_mock_inputs}, which shows the distribution of FP input parameters and their uncertainties for both the real data and the mocks. There is good agreement between the mocks and data in all cases, with the largest discrepancy in the velocity dispersion where the data exhibits a small level of skewness that is not reproduced by our assumption of a 3D Gaussian fundamental plane. Residual differences (for instance in the mean values of $r$ and $i$) can also be seen due to slight differences between the distribution of absolute magnitudes in the data and BGS mocks. Nonetheless, these do not significantly affect the ability of the mocks to reproduce the most salient features of the data such as the log-distance ratio errors or velocity clustering.
        
    \subsection{Generating mock measurements of log-distance ratios}
    \label{sec:fp_mocks:observing}
    
        We take each of the FP mocks and process them with 
        the exact same Fundamental Plane fitting pipeline as used for the data \citep{DESIPV_Ross}. 
        This takes the input FP observables and their uncertainties, finds the best-fit FP and then uses this to recover observed log-distance ratios. 
        As with the data, an iterative scheme is used to remove the effects of outliers on the FP fit and corrections for the selection effects are also applied at this stage. 
        The mean, uncertainty and skewness of the log-distance ratio for each mock galaxy is reported. 
        
\begin{figure}[t]
            \centering
            \includegraphics[width=\linewidth]{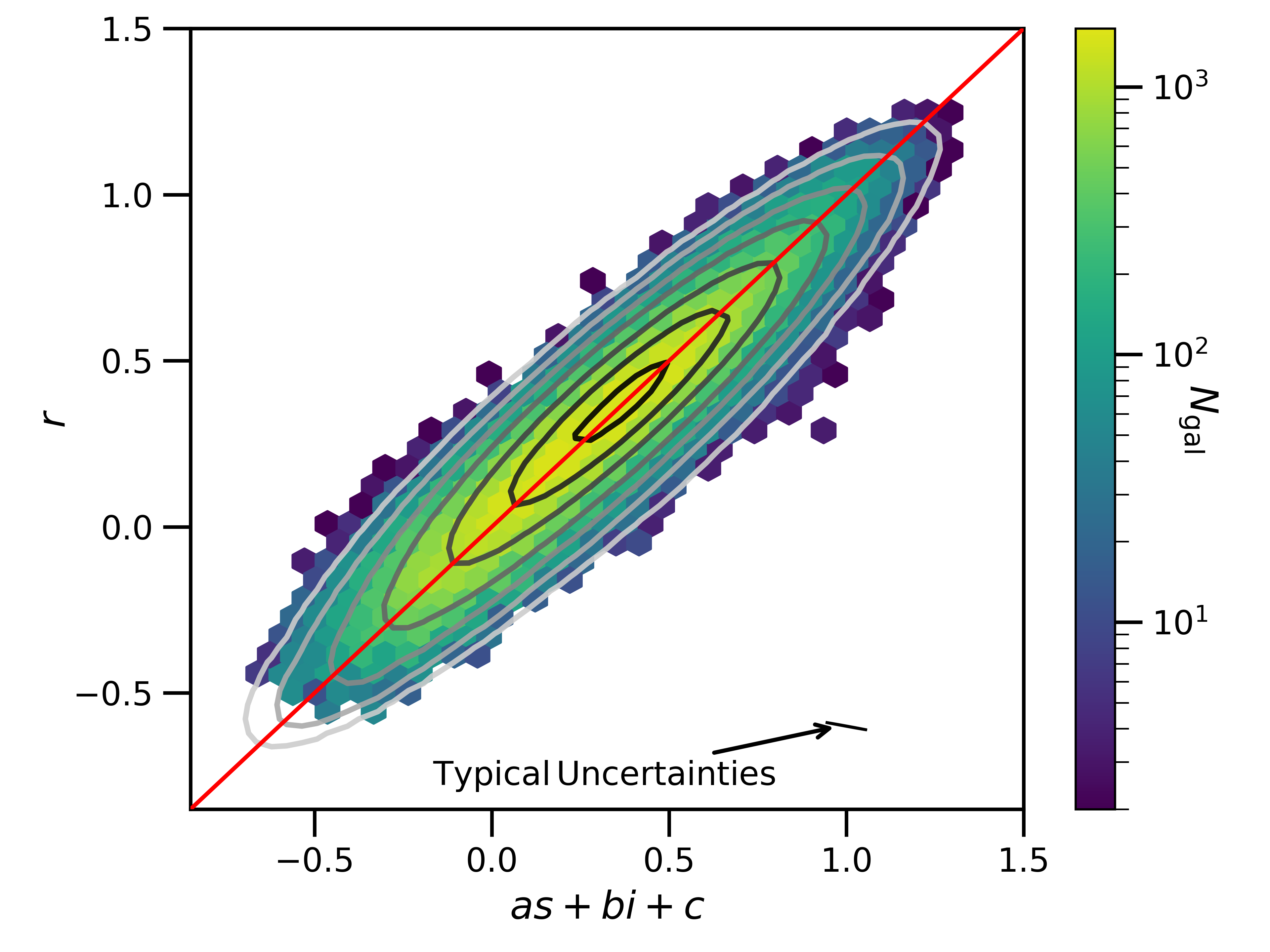}
            \caption{The fundamental plane of the data (coloured hexbins) against the average over 675 mocks (contours). The $x$-axis shows the predicted (logarithmic) effective radius of a galaxy given the surface brightness and velocity dispersion after fitting to find the best-fit coefficients $a$, $b$ and $c$. The $y$-axis shows the measured effective radius. The red line is there to guide the eye, while the small ellipse shows the typical uncertainties for a single galaxy.}
            \label{fig:fp_mock_plane}
        \end{figure}

        Figure~\ref{fig:fp_mock_plane} shows the excellent agreement between the Fundamental Plane of the data 
        alongside the average Fundamental Plane recovered for the mocks. The unique best-fit FP parameters for the data and for each mock are used to construct the $x$-axis of this plot --- however, all of the best-fit values are similar enough that using a single set of best-fit FP parameters (e.g., those of the data) for each of the mocks does not noticeably affect this plot.

\begin{figure}[t]
            \centering
            \includegraphics[width=\linewidth]{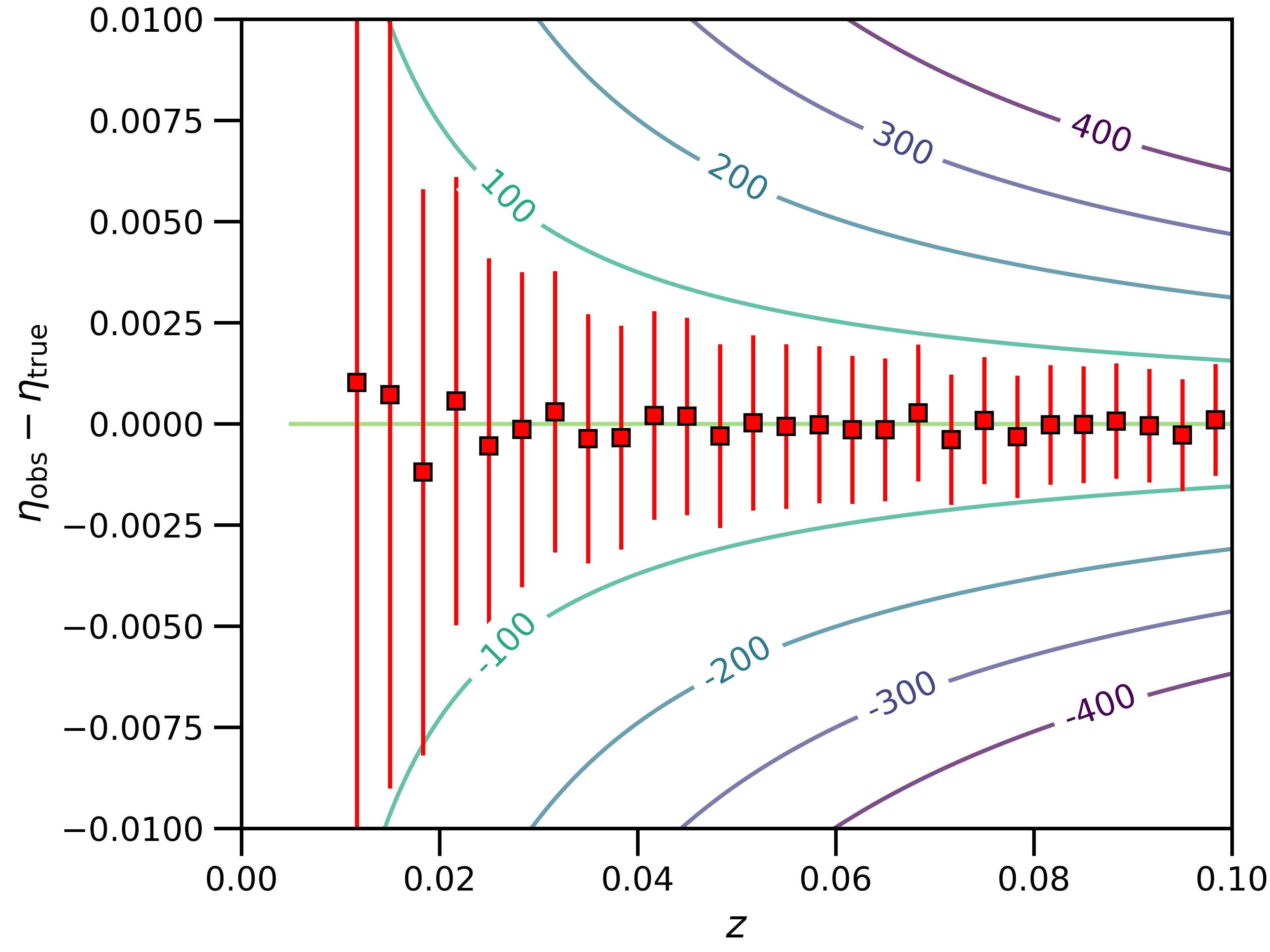}
            \caption{Residuals between observed and true log-distance ratios as a function of redshift for the Fundamental Plane mocks. 
            Each data point shows the average bias over all 675 mocks for measurements that fall in that redshift bin. The error bars represent the standard deviation over the mocks (not the error on the mean, which would be $\sqrt{675}\times$ smaller). Colored lines display values of the corresponding peculiar velocities, in units of $\mathrm{km\,s^{-1}}$.
            Negative values indicate velocities towards the observer.}
            \label{fig:fp_mock_logdist_vs_red}
        \end{figure}

      Figure~\ref{fig:fp_mock_logdist_vs_red} shows residuals between observed and true log-distance ratios for mocks as a function of redshift. 
      Across the full redshift range, mocks demonstrate excellent consistency between the measured and true log-distance ratios, with a scatter consistent with observational uncertainties. 
        
    \subsection{Gaussianizing log-distance ratios}
    \label{sec:fp_mocks:gaussianization}
          
        \begin{figure}[t]
            \centering
            \includegraphics[width=\linewidth]{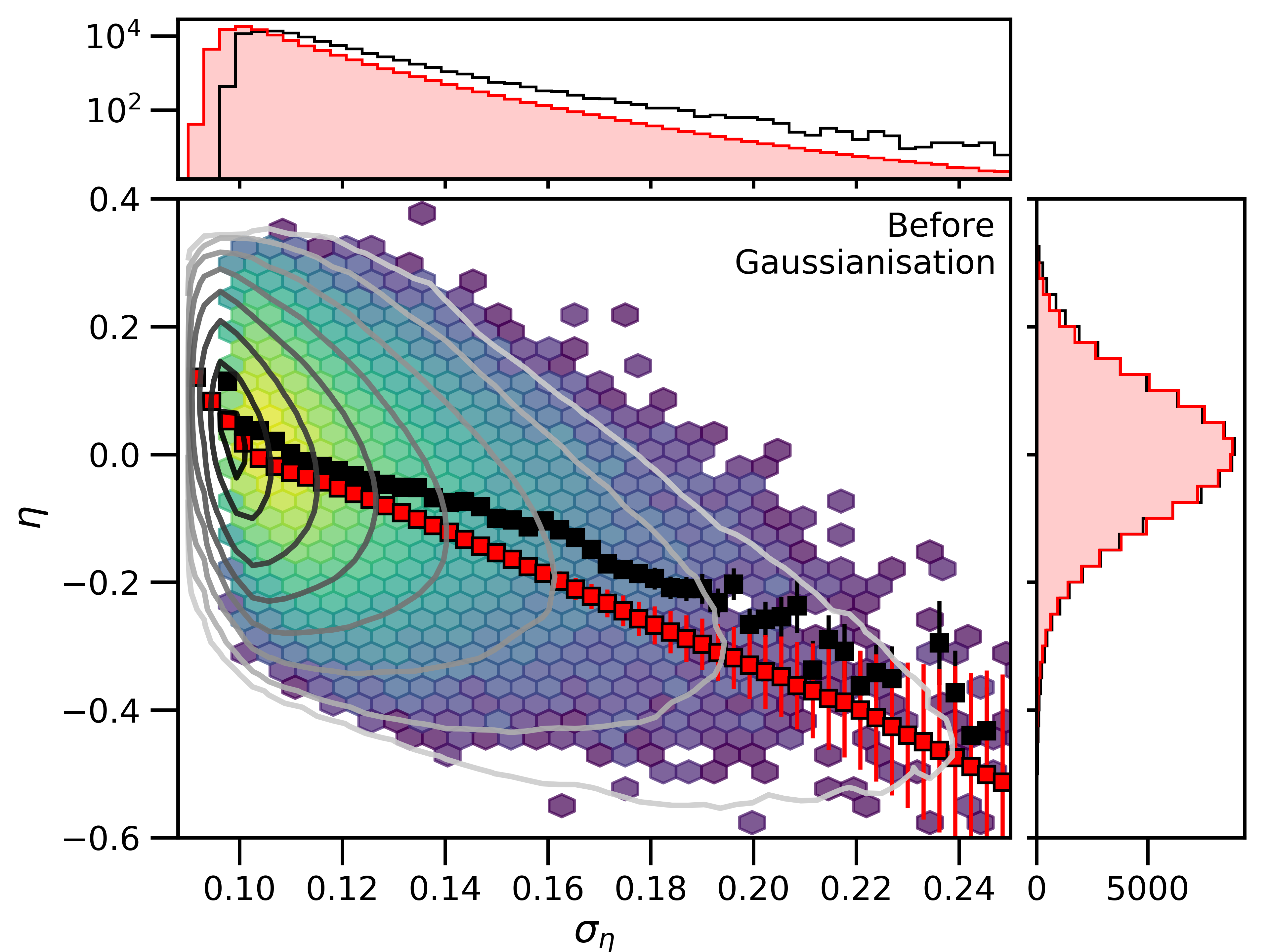}\\
            \includegraphics[width=\linewidth]{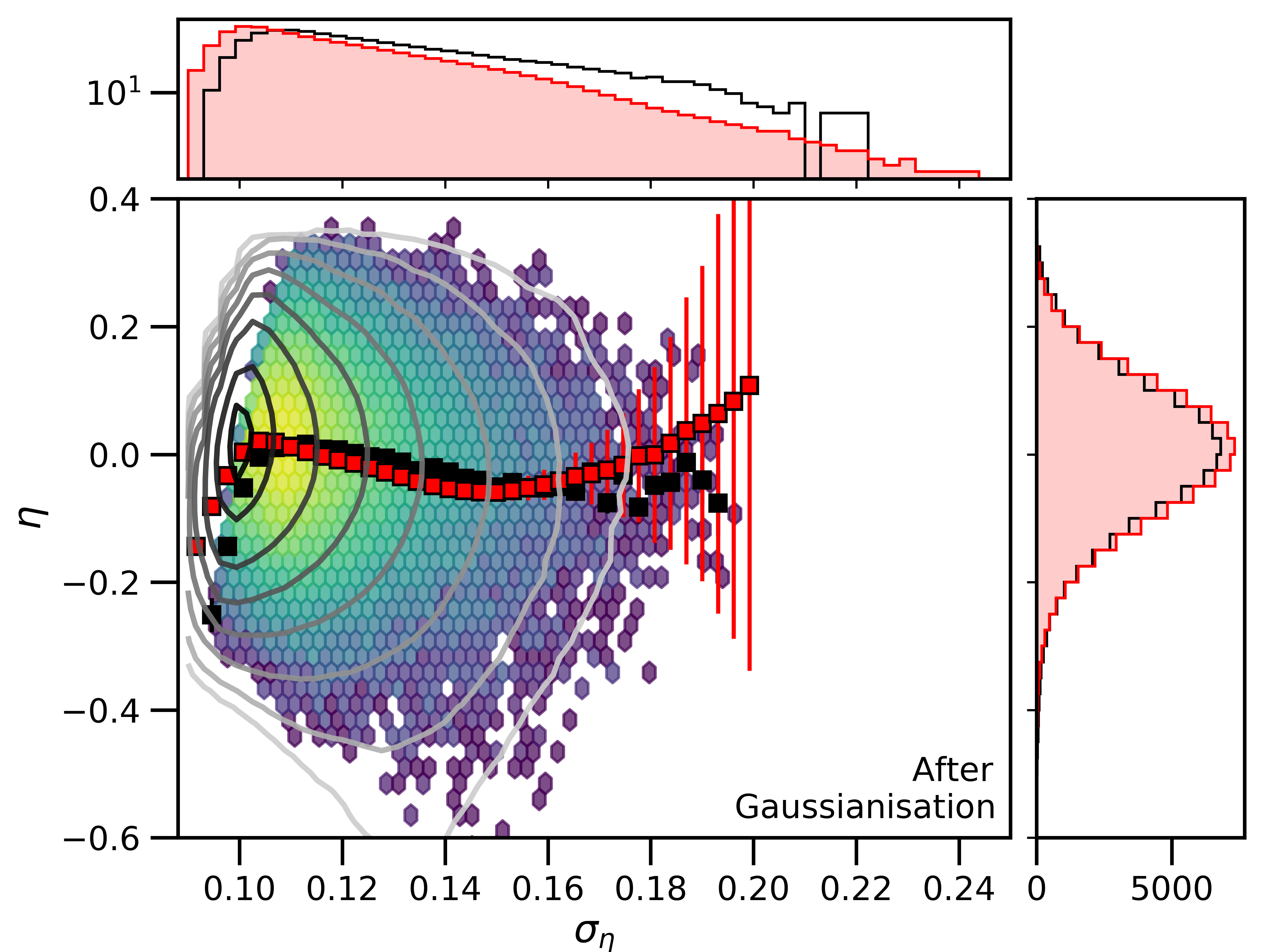}
            \caption{Observed log-distance ratios from Fundamental Plane against 
            estimated uncertainties before (top) and after (bottom) the Gaussianization procedure.
            In both cases, contours in the main panel show the mock average, with darker contours indicating regions with greater galaxy density, while the hexbins show the real DESI FP data, with brighter colors indicating more galaxies. 
            The black and red points with error bars show the mean and standard deviation of the data and mocks respectively, in bins of log-distance ratio uncertainty. The side panels show corresponding 1D histograms of the log-distances and their uncertainties for the data (black, unfilled) and mock average (red, filled).}
            \label{fig:fp_mock_logdist_vs_err}
        \end{figure}
        During preliminary analysis of the FP mocks and data, it was found that the overall distribution of log-distance ratios exhibited significant skewness towards negative log-distance ratios. 
        In the mocks, it was also found that there was a one-sided correlation between the observational uncertainty and residual (observed versus true) log-distance ratios.  Galaxies with large uncertainties were more likely to be scattered towards negative values than positive values. 
        As discussed in \citet{DESIPV_Ross}, this is mostly due to the correction for the selection function (Malmquist bias) applied to the individual log-distance ratios. 
        This is such that galaxies with negative log-distance ratios (those that would lie above the best-fit line in a Hubble diagram of distance modulus vs. redshift) are shifted even more negative (even higher in the aforementioned Hubble diagram) to account for missing galaxies, and their uncertainty increases. 
        
        The methodology ensures the inverse-variance-weighted average of the log-distance ratio remains unbiased. However, the introduction of skewness causes the weighted and \textit{un}weighted averages to differ, which can have a knock-on effect on clustering statistics that assume a Gaussian distributed set of peculiar velocities. 
        To correct for this, we apply a Gaussianization procedure 
        to the mocks (and data), described in \citet{DESIPV_Ross}. 
        This renormalizes the uncertainties for each data point such that the weighted and unweighted averages agree, while also preserving the mean and standard deviation of the uncertainties.

        As can be seen in Figure~\ref{fig:fp_mock_logdist_vs_err}, this successfully removes the correlation between the observed log-distance ratio and uncertainty while keeping the uncertainties representative of the scatter between observed and true log-distance ratios. 
        The trends in the mocks before and after correction match those in the data very well. 
        The log-distance ratio uncertainties in the data are slightly larger than the mock \textit{average} even before the Gaussianisation procedure is applied. 
        However, these uncertainties are not larger than those seen across the \textit{range} of mocks (given that some mocks have slightly better or worse measured log-distance ratios than the average), and so while this could be looked at further for future DESI data releases the mocks here are suitably representative of the data for our purposes.
    
\section{The Tully-Fisher mocks}
\label{sec:tf_mocks}
    
    The Tully-Fisher (TF) relation describes an empirical relationship between the brightness of a spiral galaxy and its rotational velocity at some fixed galactocentric radius \citep{tullyNewMethodDetermining1977}. 
    As described in \cite{douglassDESIEDRCalibrating2025} and \cite{DESIPV_Douglass}, the DESI TF sample is based on a sample of late-type galaxies from the Siena Galaxy Atlas \citep[SGA]{moustakasSienaGalaxyAtlas2023}, a size-limited catalog built from the DESI Legacy Imaging Surveys \citep{deyOverviewDESILegacy2019}. 
    For each galaxy, one DESI fiber is placed at its galactic center and a second fiber at a distance $0.4R_{26}$ from the center, where $R_{26}$ is the radius of the galaxy's 26th magnitude elliptical isophote, measured in the $r$-band. 
    The difference in redshift between the observation at the galactic center and $0.4R_{26}$ provides a measurement of the component of the galaxy's radial velocity along the line of sight.
    
    To forecast the sensitivity of the PV survey to the growth of structure, the TF mocks must reproduce the angular and magnitude selection functions of the DESI PV catalog \citep{saulderTargetSelectionDESI2023}, as well as the number density and log-distance ratio of the target galaxies. The following subsections describe the procedure to generate the mocks and checks of their validity.
    
    \subsection{Basic selection}
    \label{sec:tf_mocks:data_cuts}
    
        Construction of the TF mock catalogs begins by cross-matching the BGS mocks with real DESI BGS data using the \texttt{TARGETID}, \texttt{SURVEY}, \texttt{PROGRAM} and \texttt{HEALPIX} keywords described in Section~\ref{sec:bgs_base_mocks:properties}. 
        
        The BGS data are required to have a valid SGA ID, which enforces a size selection $>20$~arcseconds in the SGA catalog. Basic quality cuts on the output of the spectroscopic pipeline ($\Delta\chi^2>25$ and \texttt{ZWARN=0}) reproduce the same cuts applied in the TF analysis.
        As described in \cite{douglassDESIEDRCalibrating2025}, 
        we then apply four photometric corrections to the mock galaxies:
        \begin{enumerate}[itemsep=4pt]
            \item A systematic correction for the photometric offset between the BASS and MzLS targeting surveys, $A_\text{sys}$.
            \item A $k$-correction to $z=0.1$, $A_k$, using the calculation in \cite{blantonKCorrectionsFilterTransformations2007}.
            \item Global Milky Way dust corrections using the dust maps from \cite{zhouStellarReddeningMap2025a}, $A_\text{MW}$.
            \item Per-galaxy dust corrections, $A_\text{dust}$, that account for internal extinction occurring in galaxies with higher inclination angles to our line of sight.
        \end{enumerate}
        The corrections are used to adjust $m_r$, the apparent $r$-band magnitude within the 26-mag isophote of each galaxy, according to
        \begin{equation}
            m_{r, \text{corr}} = m_r - A_{\rm MW~dust} - A_{\rm internal~dust} + A_k + A_{\rm sys}.
            \label{eq:tf_syst_corr}
        \end{equation}
        
\begin{figure}[t]
            \centering
            \includegraphics[width=\linewidth]{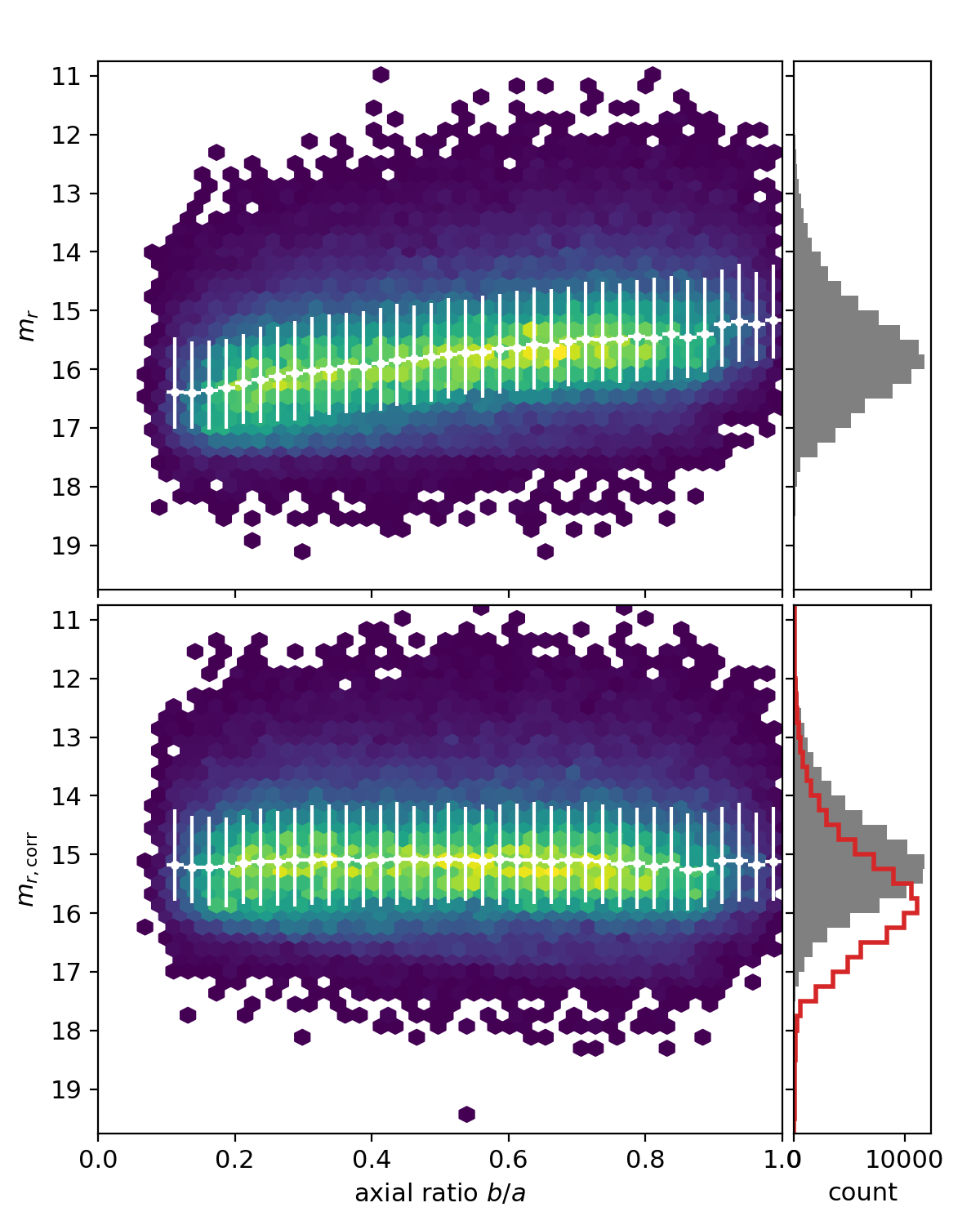}
            \caption{Apparent $r$-band magnitude at $R_{26}$ as a function of axial ratio $b/a$ for the Tully-Fisher mocks before (top) and after (bottom) the photometric corrections in Eq.~\eqref{eq:tf_syst_corr}. The white points indicate the mean $m_r$ in each bin of $b/a$, and the error bars correspond to the sample standard deviation per bin. 
            The gray histograms show the distributions of $m_r$ and $m_{r,\mathrm{corr}}$, with the uncorrected $m_r$ distribution (red line) superimposed on $m_{r,\mathrm{corr}}$ in the bottom right. }
            \label{fig:tf_mock_syst_corr}
        \end{figure}

        Figure~\ref{fig:tf_mock_syst_corr} shows how this correction 
        successfully makes average magnitudes to be independent 
        of axial ratio $b/a$. 
        
        The following late-type selection cuts defined in \cite{saulderTargetSelectionDESI2023} complete the generation of the mock sample:
        \begin{enumerate}[itemsep=4pt]
            \item Basic cuts that remove corrupt photometry 
            \item Axial ratio $b/a < \cos{25^\circ}$.
            \item Either exponential morphology or Sérsic morphology with Sérsic index $n_s\leq2$.
        \end{enumerate}
        Note that several of the cuts are  {roughly the opposite} of the early-type selection applied to produce the FP sample (section~\ref{sec:fp_mocks:data_cuts}), 
        making these samples disjoint. 
        
    \subsection{Generating mock rotational velocities}
    \label{sec:tf_mocks:noisy_tf_parameters}
    
        We simulate the TF relation by sampling from the 
        magnitude-velocity relation observed in data. 
        We begin by computing the true absolute magnitude $M_{r,\text{cos}}$ using the cosmological redshift from the mocks,
        \begin{equation}
            M_{r,\text{cos}} = m_{r,\text{corr}} - \mu(z_\text{cos}),
        \end{equation}
         {assuming a cosmology that matches the one from the \textsc{AbacusSummit} suite but with} $H_0=100$~km~s$^{-1}$~Mpc$^{-1}$. Note that $M_{r,\text{cos}}$ explicitly excludes the peculiar velocities in the mock catalogs.
        
        Using the same flat $\Lambda$CDM cosmology, we compute inferred distance moduli $\mu(z_\text{obs})$ from the redshifts $z_\text{obs}$ in the mocks, which include peculiar motion. From these, we generate the inferred absolute magnitudes of the mock galaxies:
        \begin{equation}
            M_{r,\text{obs}} = m_{r,\text{corr}} - \mu(z_\text{obs}).
        \end{equation}
        
\begin{figure}[t]
            \centering
            \includegraphics[width=\linewidth]{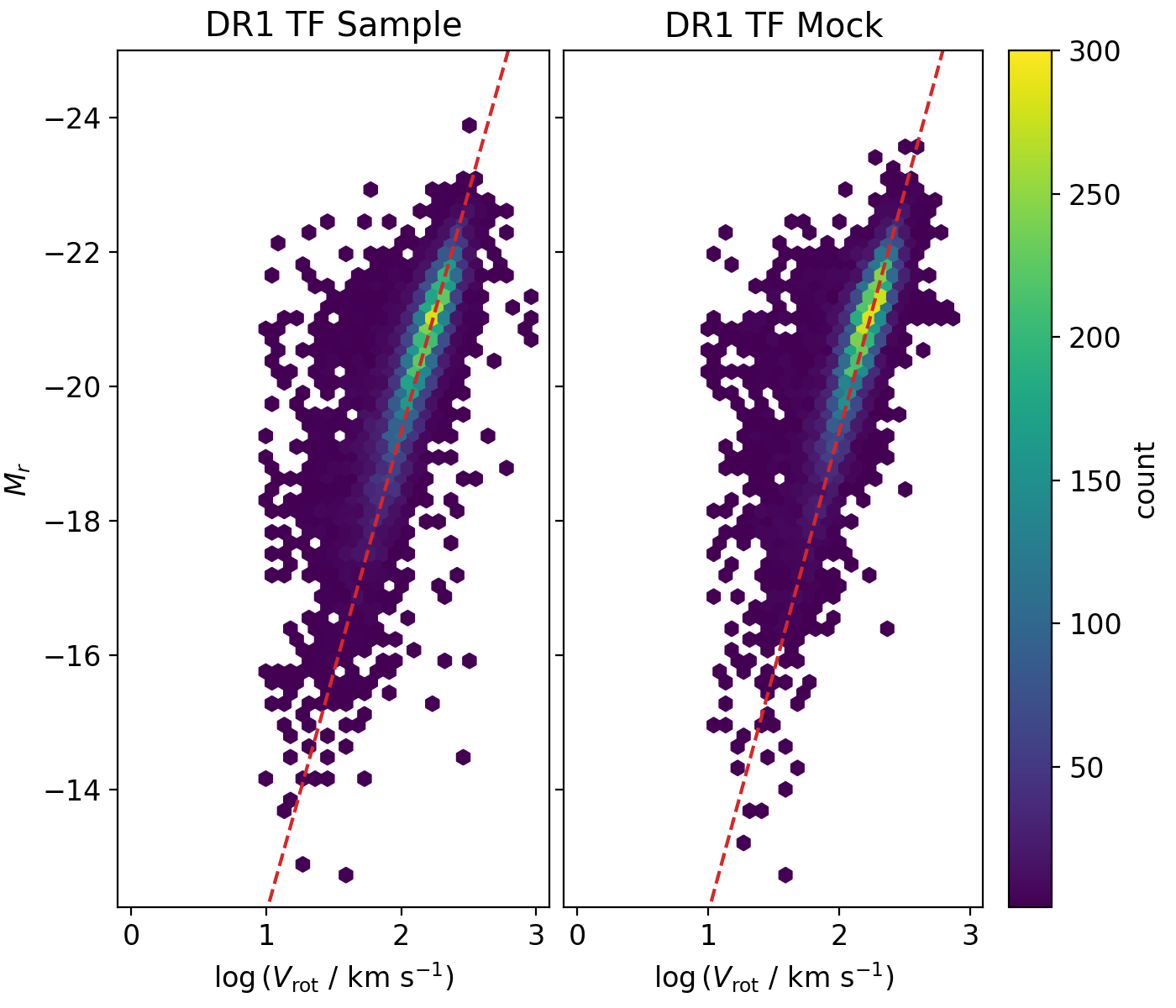}
            \caption{The DESI DR1 Tully-Fisher sample from \cite{DESIPV_Douglass} (left) and a mock TF dataset downsampled to match the statistics of the DR1 catalog (right) using the generation procedure described in the text. The dashed lines indicate the best-fit slope of the TF relation from DR1.}
            \label{fig:tf_mock_dr1_relation}
        \end{figure}

        At this stage, we produce mock rotational velocities from the data themselves. Using the measured $\log{V_\text{rot}}$ from the DESI DR1 TF sample, and the absolute magnitude $M_{r,\text{TF}}$ inferred from the best-fit TF relation, we bin the data in $M_{r,\text{TF}}$ using bins of width 0.05 mag. In each magnitude bin, we construct a smoothed cumulative distribution function (CDF) of $\log{V_\text{rot}}$ using the TF sample. Finally, we loop through the mock catalog, sorting galaxies by $M_{r,\text{cos}}$ into the bins in $M_{r,\text{TF}}$. Using the CDF of $\log{V_\text{rot}}$ in each magnitude bin, we randomly generate mock rotational velocities for all mock galaxies using inverse transform sampling.
        Figure~\ref{fig:tf_mock_dr1_relation} compares the 
        magnitude-velocity distributions between data and mock,
        showing excellent agreement. 
        
        In the final step, we assign rotational velocity uncertainties to each mock galaxy using the data themselves. We construct a $k$-d tree using $\log{V_\text{rot}}$ and $M_{r,\text{TF}}$ from the DESI DR1 catalog and, for each mock galaxy, we find the nearest neighboring data point in the space of $(\log{V_\text{rot}}, M_{r,\text{TF}})$. 
        The rotational velocity uncertainty for each mock galaxy is assigned from its corresponding nearest DR1 data point.
        
        
        
        
        
        
        
        

        
    \subsection{Generating mock measurements of log-distance ratios}
    \label{sec:tf_mocks:observing}
    
        \begin{figure}[t]
            \centering
            \includegraphics[width=\linewidth]{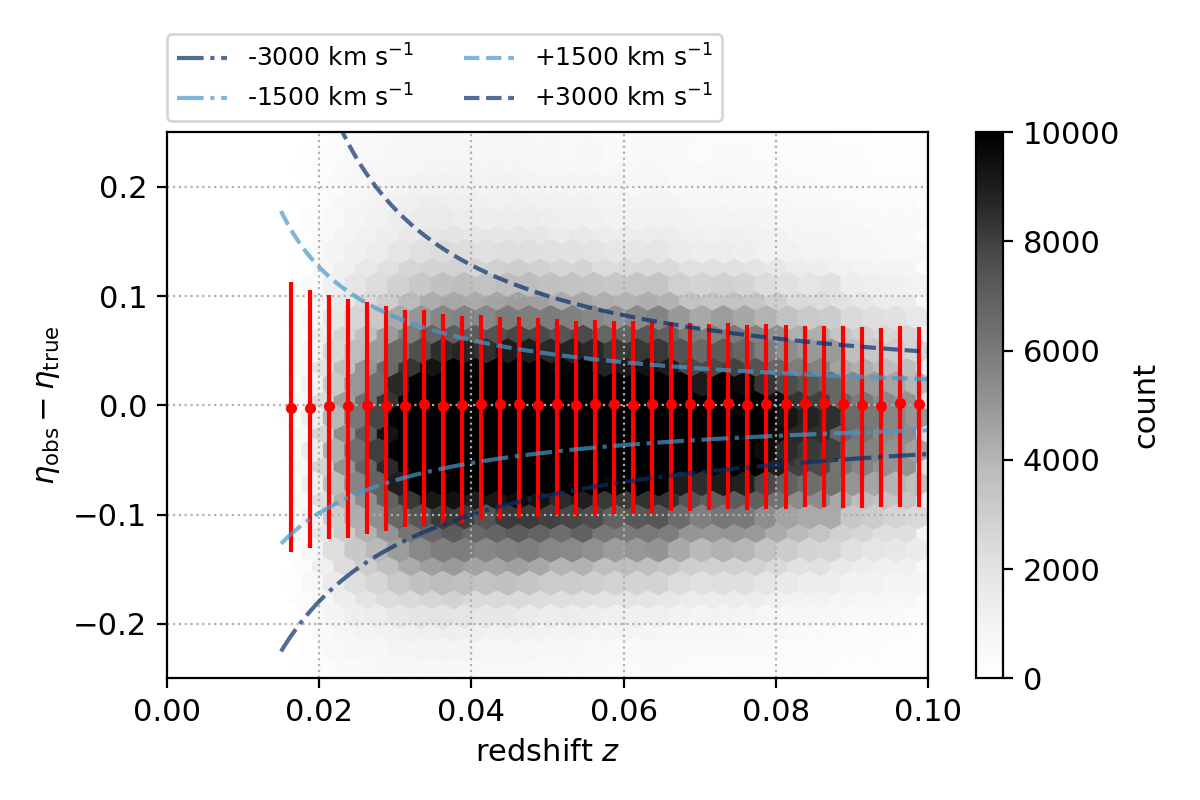}
            \caption{Residuals between observed and true log-distance ratios as a function of redshift for the Tully-Fisher mocks. As in Figure~\ref{fig:fp_mock_logdist_vs_red}, the red points show the average bias over all 675 mocks for measurements in each redshift bin, and the error bars represent the sample standard deviation in each bin. Coloured lines are lines of constant peculiar velocity. The 2D histogram indicates the underlying distribution of mock galaxies.}
            \label{fig:tf_mock_logdist_vs_redshift}
        \end{figure}
        
        To reproduce the TF fit applied to DR1, we downsample each mock catalog to match the statistics of the DR1 sample (about 10,000 galaxies) and reproduce the fitting procedure described in \cite{DESIPV_Douglass}. 
        The mock TF fit includes the same cuts in rotational velocity and magnitude applied to the data to eliminate contamination by dwarf galaxies. 
        The fit allows us to predict the absolute magnitude of each galaxy according to the mock TF relation, $M_{r,\mathrm{mock}}$. Using the value $m_{r,\mathrm{corr}}$ described in the previous section, we can compute the mock distance modulus for each galaxy:
        \begin{equation}\label{eq:tf_mock_distmod}
            \mu_\mathrm{mock} =  m_{r,\mathrm{corr}} - M_{r,\mathrm{mock}}.
        \end{equation}
        
        From this value, we may compute the logarithm of the ratio between the distance to the galaxy inferred from the Tully-Fisher relation and the luminosity distance obtained from our reference cosmology, $D_z$, referred to as ``log-distance ratio'':
        \begin{equation}
            \eta_\mathrm{mock} = \log_{10}{\left(\frac{D_z}{D_\mathrm{TF,mock}}\right)}
            = \frac{\mu(z_\mathrm{obs}) - \mu_\mathrm{mock}}{5}.
        \end{equation} 
        Analogously, the ``true'' log-distance ratio is given by
        \begin{equation}
            \eta_\mathrm{true} = \frac{\mu(z_\mathrm{obs}) - \mu(z_\mathrm{cos})}{5}.
        \end{equation}
        Figure~\ref{fig:tf_mock_logdist_vs_redshift} shows the 
        residual difference between mock and true log-distance ratios as a function of redshift. 
        We find the TF relation produces an unbiased recovery of the log-distance ratio $\eta$, but with much larger variance than the FP sample (see Figure~\ref{fig:fp_mock_logdist_vs_red}).
        Even though these are larger uncertainties than the FP sample, they do contribute statistically to our 
        final growth rate measurements. 

\section{Type-Ia supernova mocks}
\label{sec:sn_mocks}

    Type-Ia supernovae (SNIa) are excellent distance indicators, 
    superior to Fundamental Plane and Tully-Fisher in terms of 
    intrinsic or unmodeled scatter. Peculiar velocities can therefore be extracted from SNIa, and be used to measure the growth 
    rate of structures \citep{howlettMeasuringGrowthRate2017,kimTestingGravityUsing2019,hutererSpecificEffectPeculiar2020,carreresGrowthrateMeasurementTypeIa2023,rosselliForecastGrowthrateMeasurement2025a}.
    While DESI is not a supernova 
    survey, several external SN programs have interesting samples overlapping with the DESI footprint, which is a great opportunity to perform a combined growth rate measurement  \citep{DESIPV_Nguyen}. 
    Among those SN surveys, we can highlight 
    the Pantheon+ compilation \citep{broutPantheonAnalysisCosmological2022}, 
    Union \citep{rubinUnionUNITYCosmology2025},
    ATLAS, 
    and the Zwicky Transient Facility \citep{bellmZwickyTransientFacility2019,rigaultZTFSNIa2025}.
    The Dark Energy Survey SNIa program \citealt{vincenziDarkEnergySurvey2024}, 
    the Dark Energy Bedrock All-Sky Supernova Program 
    (DEBASS, \citealt{acevedoDarkEnergyBedrock2025,shermanDarkEnergyBedrock2025})
    or
    the Vera Rubin Observatory \citep{ivezicLSSTScienceDrivers2019}
    are in the Southern hemisphere and do not overlap with DESI.
    In this section, we describe the procedure to produce 
    SNIa mocks on top of mock DESI BGS galaxies. 
    
    Type-Ia supernova rates in galaxies depend on the galaxy properties, such as stellar mass (\(M_*\)), star formation rate (SFR), and metallicity \citep{sullivanRatesPropertiesType2006}. 
    We model the correlation between SN Ia rates ($R_{Ia}$) and galaxy properties
    using a two-component parametrization (the ``A+B'' model) introduced
    by \citet{mannucciSupernovaRateUnit2005}.
    The model of $R_{Ia}$
    of a galaxy consists
    of a “delayed” component, driven
    by the stellar mass of the galaxy, and a “prompt” component,
    caused by the formation of new stars. 
    The $R_{Ia}$ can be expressed as: 
    \begin{equation}
        R_{Ia} = \left[ A \times \left(\frac{M_*}{M_\odot}\right) + B \times \left(\frac{\mathrm{SFR}}{M_\odot \mathrm{yr}^{-1}}\right) \right] ~ \mathrm{SN \ yr}^{-1}.
        \label{eq:snIa rate}
    \end{equation} 
    \citet{vincenziDarkEnergySurvey2021} extended this model to account for
    correlations between host-galaxy mass $M_*$ and the stretch parameter 
    $x_1$ of SNIa light-curves \citep{smithFirstCosmologyResults2020}. 
    We leave investigation of potential effects of such extension 
    for future work. 
    
    In the current version of mocks, we opted to use 
    the best-fitting $A$ and $B$ parameters from 
    \citet{sullivanRatesPropertiesType2006}:
    \begin{equation}\label{eq:snIa rate2}
        \begin{split}
        A & =  (5.3)\times 10^{-14}; \\
        B & =  (3.9 ) \times 10^{-4}.
        \end{split}
    \end{equation}
    The number of supernovae for each galaxy is drawn from a Poisson distribution with expectation value of: 
    \begin{equation}
        \langle N_{Ia} \rangle = t \times  R_{Ia}
    \end{equation}
    where $t$ is the duration of the survey. 

    \citet{DESIPV_Nguyen} used those supernovae mocks to produce 
    simplified simulations of the Pantheon+ survey \citep{broutPantheonAnalysisCosmological2022} and detect 
    magnitude fluctuations due to large-scale structure, 
    mainly driven by peculiar velocities.

\section{Clustering catalogs}
\label{sec:clustering catalogs}

   \begin{figure*}
        \centering
        \includegraphics[width=0.48\textwidth]{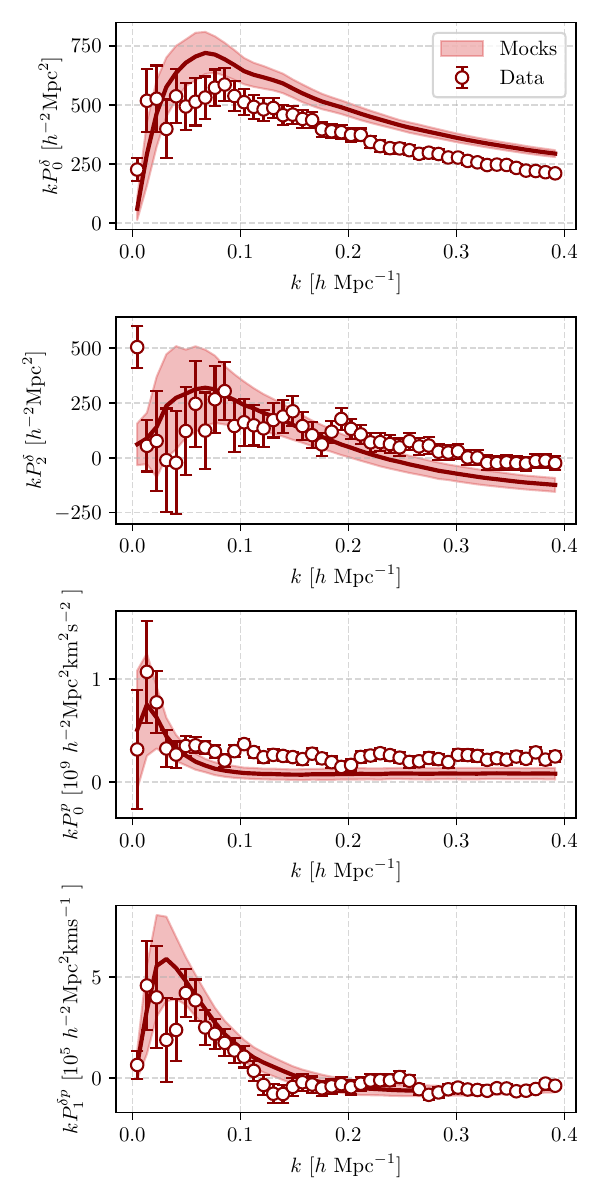}
        \includegraphics[width=0.48\textwidth]{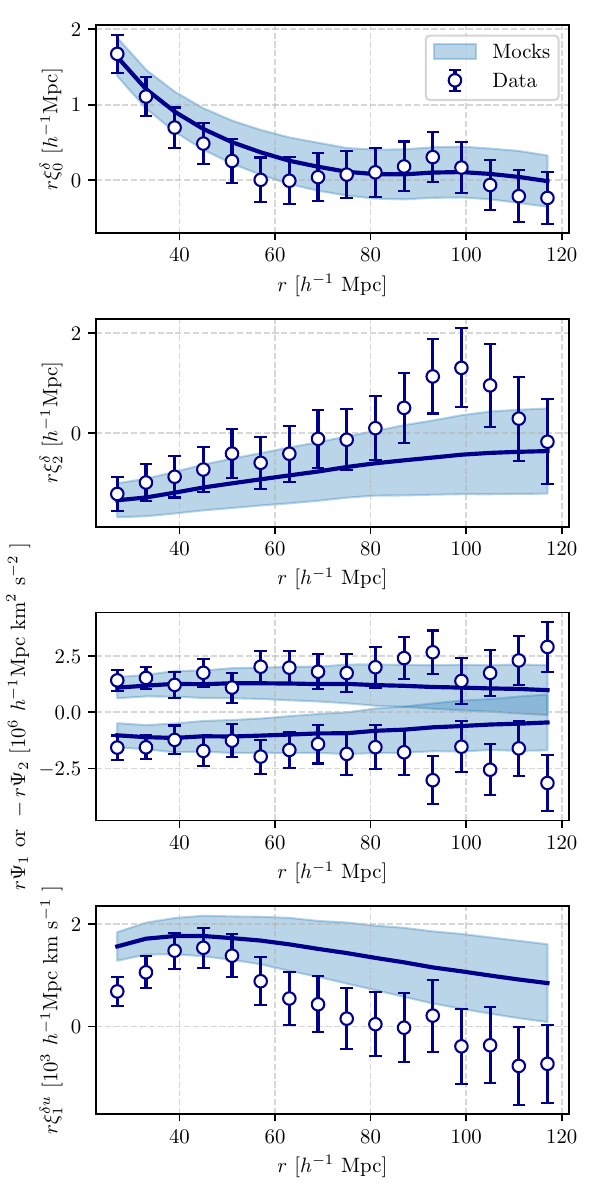}
        \caption{Two-point statistics of data versus mocks. Left panels displays Fourier space power spectra and right panels show correlation functions. 
         {The four rows show respectively the 
        monopole of the auto-correlation of galaxies, $P^{\delta}_0$ and $\xi^{\delta}_0$, the quadrupoles of the auto-correlation of galaxies, $P^{\delta}_0$ and $\xi^{\delta}_0$, the auto-correlation of momenta, $P^{p}_0$ and $\Psi_1,\Psi_2$, and the dipole of the cross-correlation between density and momenta, $P^{\delta p}_1$ and $\xi^{\delta u}_1$. }
        Shaded areas correspond to the average clustering of 675 mock realizations and their standard deviation. Data error bars are estimated from the scatter of mock measurements. Note that the effective redshift of the data is $z_\mathrm{eff}=0.07$ while for mocks we used the $z=0.2$ snapshot from the n-body simulation. }
        \label{fig:2pt_data_mocks}
    \end{figure*}
    
    From the TF and FP mocks, we construct subsets 
    of galaxy peculiar velocities that are optimal 
    for clustering measurements. 
    Those clustering catalogs are designed to retain only the most robust peculiar velocity measurements and include additional systematics weights to account for the effects of redshift completeness, fiber assignment and density variations in the imaging used for target selection \citep{desicollaborationDESI2024II2025}. 
    The additional cuts and weights are the same as for the real TF and FP catalogs; details can be found in \citet{DESIPV_Douglass,DESIPV_Ross}. 
    Following section~\ref{sec:bgs_base_mocks:randoms}, two random catalogs are also produced, with 20 and $200\times$ the number of real galaxies, matching the angular and redshift distribution of mock clustering catalogs.
    From these same random catalogs, we compute a gridded number density of BGS redshifts and peculiar velocity tracers. We then assign to each mock galaxy the corresponding value of that density at its location, which is useful 
    for calculation of optimal weights for clustering.

    Figure~\ref{fig:2pt_data_mocks} compares the two-point statistics of mocks and data, both in Fourier space (left column) and configuration space (right column).
    In Fourier space, we show the monopole and quadrupole of the auto power spectrum of galaxies $P_0^\delta$ and $P_2^\delta$, the monopole of the auto power spectrum of the momentum field $P_0^p$ and the dipole of the density momentum cross power spectrum $P_1^{\delta p}$. 
    In configuration space, we display similar statistics: monopole $\xi^\delta_0$ and quadrupole $\xi^\delta_2$ of galaxy auto-correlation function, two components of the momentum correlation functions $\Psi_1$ and $\Psi_2$, and lastly the dipole of the galaxy-momentum cross-correlation function $\xi^{\delta u}_1$. 
    Those measurements are described in \citet{DESIPV_Qin} and \citet{DESIPV_Turner} respectively. 
    The uncertainties of data points are estimated from the
    scatter of 675 mock measurements of the same statistics. 
    Note that correlation function points are highly 
    correlated. 
    The overall shape is consistent between data and mocks
    over large scale ranges. The amplitude differences 
    are mostly due to the use of the $z=0.2$ snapshot 
    to produce the mock catalogs (section~\ref{sec:bgs_base_mocks:abacus}), while the 
    data has an effective redshift of $z_\mathrm{eff}=0.07$.
    Other differences might be due to potential remaining 
    systematics on data, such as the amplitude shift seen in 
    $P^p_0$ (left column, third row). 
    
     {Differences in the clustering amplitudes of mock versus data 
    are not an issue to first order since the mock measurements only 
    impact data results through the covariance matrix. 
    The covariance matrix is a combination of cosmic variance and 
    shot noise, and the clustering amplitude mostly affects the former. 
    Therefore, an empirical scaling of the covariance matrix to account for the 
    mismatch in clustering amplitude is not simple to do. 
    We leave further investigation of this issue for future work, where 
    the precision of our uncertainties could be a more important issue. 
    }

\section{Consensus measurement for the growth rate using mocks}
\label{sec:consensus}

 {Given that we apply three different methodologies to real data, 
we aim at optimally combining the resulting $\fsig$ from those 
into a single consensus result accounting for the correlations 
between methodologies.}
Our sample of mock catalogs was used to estimate the 
correlations between our $f\sigma_8$ measurements obtained 
from those three different methodologies applied on the same data: 
momentum power spectra \citep{DESIPV_Qin}, 
correlation functions \citep{DESIPV_Turner} 
and maximum likelihood estimator \citep{DESIPV_Lai}. 

\begin{table}[t]
    \centering
    \caption{Summary statistics from the fits to 675 mocks using our three methodologies to measure $\fsig$.
    Diagonal values show the average and standard deviation for a given method, while off-diagonal terms display the estimated Pearson correlation coefficient, with uncertainties estimated with bootstrap realizations. The fiducial value of the mock is $f(z=0.2)\sigma_8(z=0.2) = 0.636 \times 0.726 = 0.462$
    .}
    \label{tab:fs8_correlations}
    
       \begin{footnotesize}

   \begin{tabular}{l c c c }
              & MLE & $P_{ij}(k)$ & $\xi_{ij}(r)$ \\
              \hline 
MLE & 0.461 $\pm$ 0.059  &  --  &  --  \\
$P_{ij}(k)$ & (38.8 $\pm$ 3.4)\%  & 0.439 $\pm$ 0.054  &  --  \\
$\xi_{ij}(r)$ & (27.3 $\pm$ 5.6)\%  & (29.0 $\pm$ 4.3)\%  & 0.478 $\pm$ 0.079  \\
    \end{tabular}
        \end{footnotesize}
    
\end{table}

\citet{sanchezClusteringGalaxiesCompleted2017a} proposed 
a method to combine $K$ Gaussian posteriors for a $d$-dimensional 
vector of parameters $\vec{p}_k$ where $k = 0, 1, \ldots, K$,
with a $d\times d$ covariance matrix 
${\bf C}_p^k$ describing the uncertainty on $\vec{p}^k$ from method $k$. 
The main idea of the method is to create a larger $Kd \times Kd$ 
parameter covariance matrix that includes cross-correlations between 
all different methods. 
To obtain such a covariance, we applied all our methodologies to 
all mock catalogs and use all best-fit parameters to obtain a sample matrix of correlation coefficients.
This method was recently employed to combine clustering analyses 
performed in configuration and Fourier space \citep{bautistaCompletedSDSSIVExtended2021,dumerchatBaryonAcousticOscillations2022} and the code implementing it is publicly available.\footnote{\url{https://github.com/TyannDB/Gacomb/}}
The final consensus has a Gaussian posterior by construction. 

In our case, we are interested in the consensus value for a single parameter $\vec{p} = \{f\sigma_8\}$ ($d=1$), and three methods ($K=3$): the maximum likelihood estimator (MLE), the  power spectra ($P_{j}(k)$) and the correlation functions $\xi_{ij}(r)$. 
Table~\ref{tab:fs8_correlations} displays the statistics of 
measurements in mock catalogs, including average $f\sigma_8$, 
the scatter over 675 realizations and the correlation coefficients
between our three methods. The uncertainties on the correlation coefficients are derived from the scatter of $10^6$ bootstrap realizations
of mock measurements. 
First, we see that the average $\fsig$ for each method is consistent with the 
fiducial value of $f\sigma_8(z=0.2) = 0.462$ within one standard deviation, which
is a proxy for the uncertainty with real data. 
The standard deviation of $\fsig$ values is also consistent between methods, 
with a slightly higher scatter for the correlation functions $\xi_{ij}$. 
Correlations among methods lie between 27.3\% and 38.8\%, since they use the same
initial dataset and measure the same parameter. 
 {The observed correlations are lower than expected. 
For instance in past full-shape analyses, Fourier and configuration space results on the growth rate are correlated at 84\% \citep{bautistaCompletedSDSSIVExtended2021}.
The lower correlations seen here might be due 
to different analysis choices, scale ranges 
and model parametrization for each method, or the lower overall signal-to-noise ratio of our measurement.  } 

Our  {real data} results from the DESI DR1 PV peculiar velocity surveys are:
\begin{itemize}[itemsep=4pt]
    \item MLE \citep{DESIPV_Lai}: $f\sigma_8 = 0.483_{-0.043}^{+0.080}(\mathrm{stat}) \pm 0.018(\mathrm{sys})$
    \item $P_{ij}(k)$ \citep{DESIPV_Qin}: $f\sigma_8 = 0.440_{-0.096}^{+0.080}$(stat)
    \item $\xi_{ij}(r)$ \citep{DESIPV_Turner}: $f\sigma_8 = 0.391_{-0.080}^{+0.081}$(stat)
\end{itemize}

 {Note that the statistical uncertainty from the MLE method has been scaled up based on results from mocks. 
For the consensus calculation, we need Gaussian symmetric uncertainties, so we use half of the confidence interval for each measurement. The systematic uncertainty has been added in quadrature to the scaled statistical uncertainty.}
Using those values and the mock correlation matrix, we obtain the following consensus result with real data:
\begin{equation}
    \mathrm{\bf DESI~DR1~PV~consensus:} \quad  f\sigma_8 = 0.450 \pm 0.055 
    \label{eq:fs8_consensus}
\end{equation}

This 12\% precision measurement is consistent with General Relativity and $\Lambda$CDM predictions within one sigma.

\section{Conclusions}

This work presented the official set of mock catalogs of the DESI DR1 Peculiar Velocity Survey. 
A set of 675 realizations of DR1 samples were produced from the \textsc{AbacusSummit} suite of n-body simulations, with galaxies added
with a nested-HOD model, reproducing the Bright Galaxy Survey number density and clustering. 
From those BGS mocks, we derived density field subsamples 
as well 
as mocks for three distance indicators: Fundamental Plane, Tully-Fisher and type-Ia supernovae. Those include 
realistic sampling of real data properties, and 
mimic real measurements of FP and TF relations to derive
peculiar velocities, including measurement uncertainties. 
Since they are constructed from the same initial simulation, 
they automatically include all cross-correlations between 
densities and peculiar velocities. 
This work delivers the largest and most realistic sample of mock catalogs for cosmology with peculiar velocities to date. 

Our mock catalogs were consistently used to test three
methodologies that yield estimates for the growth rate of 
structures \citep{DESIPV_Lai,DESIPV_Qin,DESIPV_Turner}.
Using measurements on all 675 mocks, we could derive 
the correlation between methodologies and compute 
a consensus result with the DESI DR1 sample.

Growth rate measurements will become a DESI Collaboration Key Paper with the future DR2 and DR3 datasets, and will require careful estimation of statistical and systematic uncertainties. A robust set of realistic mock catalogs will be essential to test our methodologies. 

 {The matching between data and mocks is not perfect due to several reasons: the redshift of the snapshot used from the initial simulation, the redshift range of BGS HOD fits, and potential remaining systematics on real data. However, our mock catalogs served as a self-consistency test, i.e., to check that different methodologies recover the input cosmology of the mocks, regardless of the real data. In analysis of real data, the only information coming from mocks is the covariance matrix of power spectrum and correlation function measurements, so differences in clustering amplitudes and noise properties, which are hard to correct empirically, will cause only mild changes to final uncertainties on the growth rate.}
Some approximations and hypotheses used in this work could 
be improved in future work. Among those we can cite:
\begin{itemize}[itemsep=4pt]
    \item use the $z=0.1$ snapshot of \textsc{AbacusSummit}, instead of $z=0.2$, and refit the HOD model corresponding to the correct data redshift range $z<0.1$, instead of $0.1<z<0.5$, also using the larger DR2 sample;
    \item construct mocks with different underlying cosmologies, currently available in the \textsc{AbacusSummit} suite,  {in order to check, e.g., systematic effects related to the choice of fiducial cosmology.}
    \item create mocks from a higher resolution suite of n-body simulations (e.g., OuterRim, \citealt{ishiyamaUchuuSimulationsData2021}) or different galaxy-halo connection models;
    \item create a larger number of approximate mocks for covariance matrices and statistical tests;
    \item check for the impact of additional correlations of SNIa rates and other parameters.
\end{itemize}

Those mock catalogs will be key to improving our constraints
on dark energy and modified gravity theories, since our 
growth rate measurements tackle the low-redshift epoch
when the expansion of the Universe is accelerated. 

\section*{Data availability}

The data used to produce all figures is available in \textsc{Zenodo}.\footnote{\url{https://zenodo.org/records/17349494}}

\begin{acknowledgements}
    The project leading to this publication has received funding from 
    Excellence Initiative of Aix-Marseille University - A*MIDEX, 
    a French ``Investissements d'Avenir'' program (AMX-20-CE-02 - DARKUNI).

    This material is based upon work supported by the U.S. Department of Energy (DOE), Office of Science, Office of High-Energy Physics, under Contract No. DE–AC02–05CH11231, and by the National Energy Research Scientific Computing Center, a DOE Office of Science User Facility under the same contract. Additional support for DESI was provided by the U.S. National Science Foundation (NSF), Division of Astronomical Sciences under Contract No. AST-0950945 to the NSF’s National Optical-Infrared Astronomy Research Laboratory; the Science and Technology Facilities Council of the United Kingdom; the Gordon and Betty Moore Foundation; the Heising-Simons Foundation; the French Alternative Energies and Atomic Energy Commission (CEA); the National Council of Humanities, Science and Technology of Mexico (CONAHCYT); the Ministry of Science, Innovation and Universities of Spain (MICIU/AEI/10.13039/501100011033), and by the DESI Member Institutions: \url{https://www.desi.lbl.gov/collaborating-institutions}. Any opinions, findings, and conclusions or recommendations expressed in this material are those of the author(s) and do not necessarily reflect the views of the U. S. National Science Foundation, the U. S. Department of Energy, or any of the listed funding agencies.

    The authors are honored to be permitted to conduct scientific research on I'oligam Du'ag (Kitt Peak), a mountain with particular significance to the Tohono O’odham Nation.
\end{acknowledgements}

\bibliographystyle{aa}
\bibliography{MyLibrary, FQinRef.bib, DESI_DR1_PV.bib}



\end{document}